\documentclass{article}
\usepackage{float}
\usepackage{authblk}
\usepackage{algorithm}
\usepackage{algorithmicx}  
\usepackage{algpseudocode}  
\usepackage{setspace}
\usepackage{bm}
\usepackage{url}
\usepackage{amsmath}
\usepackage{mathrsfs}
\usepackage{dcolumn}
\usepackage{graphics}
\usepackage{graphicx}
\usepackage{geometry}
\usepackage{amsfonts} 
\usepackage{amsthm}
\usepackage{subfigure}  
\usepackage{multirow}

\title{\textbf{Recursive Regret Matching: A General Method for Solving Time-invariant Nonlinear Zero-sum Differential Games}}

\author[a,b]{Wei Liao}
\author[a,b]{Xiaohui Wei \thanks{Corresponding author}}
\author[c]{Jizhou Lai}

\affil[a]{\footnotesize{ Key laboratory of Fundamental Science for National Defense-Advanced Design Technology of Flight Vehicle, 
Nanjing University of Aeronautics and Astronautics, Nanjing, Jiangsu, China, e-mail: liaowei1991@nuaa.edu.cn}}
\affil[b]{\footnotesize{State Key Laboratory of Mechanics and Control of Mechanical Structures, 
Nanjing University of Aeronautics and Astronautics, Nanjing, Jiangsu, China, e-mail: wei\_xiaohui@nuaa.edu.cn}}
\affil[c]{\footnotesize{College of Automation Engineering, Nanjing University of Aeronautics and Astronautics, Nanjing 210001, China, 
e-mail: laijz@nuaa.edu.cn}}
\date{}
\begin{document}
\maketitle

\begin{abstract}
	In this paper, a new method is proposed to compute the rolling Nash equilibrium of the 
	time-invariant nonlinear two-person zero-sum differential games. 
	The idea is to discretize the time to transform a differential game into a sequential game with several steps, 
	and by introducing state-value function, transform the sequential game into a recursion consisting of several normal-form games,
	finally, each normal-form game is solved with action abstraction and regret matching. 
	To improve the real-time property of the proposed method, the state-value function can be kept in memory.
	This method can deal with the situations that the saddle point exists or does not exist, 
	and the analysises of the existence of the saddle point can be avoided.
	If the saddle point does not exist, the mixed optimal
	control pair can be obtained. At the end of this paper, some examples are taken to illustrate the validity of the proposed method.
\end{abstract}
\quad \\
\textbf{Keywards:} Differential game, State-value function, Regret matching, Nash equilibrium.

\section{Introduction}
For the past few years, zero-sum differential game theory 
has been extensively used in decision making problems. 
Numerous method, such as gradient-based method \cite{a1}, method based on Hamilton-Jacobi equations \cite{a2} \cite{a27}, 
dynamic programming \cite{a3} \cite{a4} and reinforcement learning \cite{a8} \cite{a9}
are proposed 
to obtain some form of optimality, especially, the saddle point. 
In most researches, the existence
of the saddle point is supposed before obtaining the saddle point \cite{a5} \cite{a6} \cite{a7} \cite{a29} \cite{a30}.
Yet the reality is that the existing conditions of the saddle point 
are too harsh to satisfy. Therefore, many applications of the zero-sum
differential games are limited to linear systems \cite{a10} \cite{a11} \cite{a12}.
In addition, for a zero-sum differential game,  
the saddle point does not exist always means that the
optimal solution (Nash equilibrium solution) of the game is a mixed solution \cite{a13}.
And the mixed optimal solution is hardly obtained
once the control schemes are determined.
Therefore, how to obtain the pure or mixed optimal solution 
without the priori hypothesis of the existence of the saddle point 
is a significant research topic. This is the motive of our research. 

\par regret matching \cite{a28} \cite{a15} is a numerical method to compute
the Nash equilibrium strategy of a normal-form game. In 
regret matching framework, computers may
use regrets of past game choices to inform future choices through self-simulated play.
In every time of self-simulated play, each player selects an action 
at random with a distribution that is proportional to positive regrets which 
indicate the level of relative losses one has experienced for not having selected the action in the past \cite{a14} \cite{a23} \cite{a31}.
Over time, the average over the strategies taken in all times of self-simulated play converges to a Nash equilibrium \cite{a15}.

% \par State-value function is a powerful tool for solving dynamic programming problem. 
% The state-value function of a state is the system’s expected total reward (performance index) with respect to a control policy. 

\par In this paper, Combining regret matching with state-value function, 
we propose a new numerical method called recursive regret matching
for the zero-sum differential games with infinite time horizon. In short,
at time $t$, 
the proposed method aims to compute the optimal control 
policies of both players for a finite time horizon in the future: $[t,t+T]$. 
After being discretized in time,
with the aid of state-value function, the differential game 
is transformed as a recursion consisting of several normal-form games, each normal-form game 
is solved with action abstraction and regret matching. 
The state-value function can be stored in memory to improve the real-time property.
In addition, this method
is effective both for the situations that the saddle point exists
or does not exist. 
For the former situation, the analysises of the existence of the saddle point are unnecessary. 
For the latter situation, the mixed optimal control policy can be obtained. 
Furthermore, the proposed method has a high real-time property, it can generate 
the optimal control input according the system state in short time. And compared to 
the existing researches, our method has fewer requirements for the form of the dynamic system.

\par The remainder of this paper is organized as follows. 
Section II presents a description of the problem.
Section III gives the detailed steps of our method.  
Some numerical examples are given in
Section IV. The results are summarized in Section V.

\section{Problem Formulation}%加入纳什均衡状态值函数的定义
\par Consider the following two-person zero-sum differential game with infinite time horizon.
The system is described by the following continuous-time nonlinear equation:
\begin{align}
	\dot{x}=f(x,a,b)
\end{align}
where $x\in \mathbb{R}^n$, $a\in \mathcal{A} \subset \mathbb{R}^{m_1}$ is the input for player I, $b\in \mathcal{B} \subset \mathbb{R}^{m_2}$ 
is the input for player II, 
$f(.,.,.)$ is a smooth function. 
Let $\mathscr{A}_{t_0}^{t_1}$ and $\mathscr{B}_{t_0}^{t_1}$ denote the set of functions from the interval $[t_0,t_1]$ 
to $\mathcal{A}$ and $\mathcal{B}$ respectively. Then,
given the initial state $x_0$ and the inputs $a(.)\in \mathscr{A}_{t_0}^{t_1} $ and $b(.)\in \mathscr{B}_{t_0}^{t_1} $, 
the evolution of system (1) is represented as $\phi_{t_0}^{t_1}(.,x_0,a(.),b(.)):[t_0,t_1] \to \mathbb{R}^n$ 
and $\phi_{t_0}^{t_1}(t_0,x_0,a(.),b(.))=x_0$.
The performance index function over the time interval $[t_0,t_1]$ is
\begin{align}
	J(x_0,t_0,t_1,a(.),b(.)) =
	\int_{t_0}^{ t_1} l\left(\phi_{t_0}^{t_1}(t,x_0,a(.),b(.)),a(t),b(t)\right) \ dt
\end{align}
where $l(.,.,.)$ is a smooth function, represents the running payoff. Suppose
$a(.)$ is chosen to maximize the performance index $J(x_0,0,\infty,a(.),b(.))$ while $b(.)$ is chosen to minimize it. 

Due to the difficulties of computing the optimal control policies for an infinite time horizon, 
the rolling optimization is adopted in this paper \cite{a20}. Briefly, the basic idea is, 
at any time $t$, to design "open-loop Nash equilibrium control inputs" 
within a moving time frame located at time $t$, regarding $x(t)$ as the initial condition
of a state trajectory $ \hat{\phi}_0^{T}(.,x(t),\hat{a}(.),\hat{b}(.)) $, 
where $\hat{a}(.) \in \mathscr{A}_0^{T}$, $\hat{b}(.) \in \mathscr{B}_0^{T}$, $T$ is the predictive period.
To distinguish them from the real variables, the hatted variables are defined as
the variables in the moving time frame.
The actual control inputs $a(t)$ and $b(t)$ 
are given by the initial value of the optimal control
inputs $\hat{a}^*(.)$ and $\hat{b}^*(.)$. That is 
\begin{align}
	a(t)=\hat{a}^*(\tau),b(t)=\hat{b}^*(\tau) \text{ when } \tau =0
\end{align}
Let the upper and lower state-value functions in the moving time frame with predictive period $T$ be defined as
\begin{align}
	\begin{split}
		\hat{V}_T^u(x_0)=\min_{\hat{b}\in \mathscr{B}_0^T }\max_{\hat{a}\in \mathscr{A}_0^T}J(x_0,0,T,\hat{a}(.),\hat{b}(.))\\
		\hat{V}_T^l(x_0)=\max_{\hat{a}\in \mathscr{A}_0^T }\min_{\hat{b}\in \mathscr{B}_0^T }J(x_0,0,T,\hat{a}(.),\hat{b}(.))
	\end{split}
\end{align}
respectively.
Obviously, $ \hat{V}_T^u(x_0)\geq \hat{V}_T^l(x_0) $ \cite{a16} \cite{a17}. 

If $ \hat{V}_T^u(x_0)= \hat{V}_T^l(x_0) $ holds for any $x_0\in \mathbb{R}^n$,
we say that the saddle point exists and the corresponding
optimal (Nash equilibrium) control pair is a deterministic solution denoted by $(\hat{a}^*(.),\hat{b}^*(.))$. 
That means under the optimal control policy, 
both players choose a single action with probability $1$ at any state and time.

If the saddle point does not exist, things will get complicated. The 
optimal control pair is no longer a deterministic solution 
but a mixed solution. That means under the optimal control pair, 
both players have at least two actions that are played with positive
probability at some states and time. 
The goal of this paper is 
to find the optimal control pair in the moving time frame
for the situations that the saddle point exists or does not exist.

\section{Method Details}
\subsection{The upper and lower state-value functions}
\par Firstly, we introduce the method to compute 
the upper and lower state-value functions. 
% For the sake of simplicity,  
% we introduce the translation operator $ \mathscr{T}_{\tau}$
% that can translate a function of time along time axis by distance $\tau$. 
% That is, for any function of time $g(.)$, 
% \begin{align}
% 	\mathscr{T}_{\tau}g(t)=g(t-\tau)
% \end{align}
Due to the Markov property \cite{a18} \cite{a19},
\begin{align}
	\begin{split}
	\hat{V}_T^u(x_0)&=\min_{b\in \mathscr{B}_0^\tau}\max_{a\in \mathscr{A}_0^\tau} 
	\bigg[J(x_0,0,\tau,a(.),b(.))+  
	 \min_{b'\in \mathscr{B}_\tau^T}\max_{a'\in \mathscr{A}_\tau^T}  J(x_0',\tau,T,a'(.),b'(.))\bigg]\\
	&=\min_{b\in \mathscr{B}_0^\tau}\max_{a\in \mathscr{A}_0^\tau} 
	\bigg[J(x_0,0,\tau,a(.),b(.))+\hat{V}_{T-\tau}^u(x'_0)\bigg]  
    \hat{V}_T^l(x_0)\\
    &=\max_{a\in \mathscr{A}_0^\tau} \min_{b\in \mathscr{B}_0^\tau}
	\bigg[J(x_0,0,\tau,a(.),b(.))  
    + \max_{a'\in \mathscr{A}_\tau^T} \min_{b'\in \mathscr{B}_\tau^T} J(x_0',\tau,T,a'(.),b'(.))\bigg]\\
    &=\max_{a\in \mathscr{A}_0^\tau} \min_{b\in \mathscr{B}_0^\tau}\bigg[J(x_0,0,\tau,a(.),b(.))+\hat{V}_{T-\tau}^l(x'_0)\bigg]
	\end{split}	
\end{align}
Here, $x_0'=\phi_0^\tau(\tau,x_0,a(.),b(.))$ and $0<\tau<T$. 
We discretize the time into $h$ intervals with size $\Delta t=\frac{T}{h}$. 
If $\Delta t$ is small enough, 
$a(.)$ and $b(.)$ can be regarded as a 
constant in interval $[k\Delta t, (k+1)\Delta t)$,
and the system (1) can be 
converted into the discretized form:
\begin{align}
	x(t+\Delta t)=x(t)+f(x(t),a(t),b(t))\Delta t \nonumber
	=F(x(t),a(t),b(t))
    % s(t+\Delta t)=s(t)+\hat{f}(s(t),u_p(t),u_e(t))\Delta t+\frac{1}{2} \frac{ \partial \hat{f} }{ \partial s }
    % \hat{f}(s(t),u_p(t),u_e(t))\Delta t^2=\hat{F}(s(t),u_p(t),u_e(t) )
\end{align}
Then 
\begin{align}
	\begin{split}
		&\hat{V}_{\Delta t}^u(x_0)=\min_{b\in \mathcal{B}} \max_{a\in\mathcal{A}}  l\left(x_0,a,b   \right)\Delta t \\
		&\hat{V}_{k\Delta t}^u(x_0)=\min_{b\in \mathcal{B}} \max_{a\in\mathcal{A}} \bigg[ l\left(x_0,a,b   \right)\Delta t
		   +\hat{V}_{(k-1)\Delta t}^u\left(F(x_0,a,b)\right) \bigg]\\
		   &\hat{V}_{\Delta t}^l(x_0)=\max_{a\in\mathcal{A}}  \min_{b\in \mathcal{B}}   l\left(x_0,a,b   \right)\Delta t \\
		   &\hat{V}_{k\Delta t}^l(x_0)=\max_{a\in\mathcal{A}} \min_{b\in \mathcal{B}}  \bigg[ l\left(x_0,a,b   \right)\Delta t
			  +\hat{V}_{(k-1)\Delta t}^l\left(F(x_0,a,b)\right) \bigg]\\
	\end{split}
\end{align}
% for $k=2,...,h$.

Equation (7) reveals a recursive form of the upper and lower state-value functions. 
With the aid of interpolation, we can obtain the 
upper and lower state-value functions, $\hat{V}_{m\Delta t}^u(.)$ and $\hat{V}_{m\Delta t}^l(.)$ 
for $m=1,2,...$, via $m$-step recursion.
see Algorithm 1
(For the sake of brevity, we assume that system (1) is two-dimensional, and denote the system state as $x=(\alpha,\beta)^\mathrm{T}$).
In Algorithm 1, 
$\widehat{V}(.)$ is a linear interpolation function 
using array $\mathcal{V}$ (the interpolation method used in this paper is shown in Appendix), and $x'=F(x,a,b)$. 
$\hat{V}_{T}^u(.)$ and $\hat{V}_{T}^l(.)$ can be obtained with input $m=h$.
\begin{algorithm}[h]
    \caption{Computation of upper and lower state-value functions}
	\begin{algorithmic}[1]
		\State \textbf{Input:} Recursion number $m$;
	  \State Select a proper rectangular computational domain $\Omega=[\alpha_l,\alpha_u]\times [\beta_l,\beta_u]$
	  (In engineering practice, we can make a rough analysis about the system possible system states to determine the computational domain); 
      \State Discretize $\Omega$ into a $N_\alpha\times N_\beta $ Cartesian grid structure;
	  \State Let $\mathcal{V}$ and $\mathcal{V}'$ be two $N_\alpha\times N_\beta$ arrays and 
	  initialize them to $\mathbf{0}$;
    %   and let $\mathcal{V}[i_\alpha][i_\beta]$ denote the state-value of state 
	%   $\displaystyle{x_{i_\alpha,i_\beta}=\left(\alpha_l+\frac{ \alpha_u-\alpha_l }{ N_\alpha-1 }i_\alpha,
	%   \beta_l+\frac{ \beta_u-\beta_l }{ N_\beta-1 }i_\beta\right)^\mathrm{T}}$; 
    %   \For{$i_\alpha\leftarrow 0,...,N_\alpha-1 $}
    %     \For{$i_\beta\leftarrow 0,...,N_\beta-1 $}
    %         \State $\mathcal{V}[i_\alpha][i_\beta]\leftarrow 0$;
    %     \EndFor
    %   \EndFor
    %   \State \quad
      \For{$k \leftarrow 1,...,m$}
        \For{$i_\alpha\leftarrow 0,...,N_\alpha-1 $}
		  \For{$i_\beta\leftarrow 0,...,N_\beta-1 $}
		  	\State $x\leftarrow \displaystyle{\left(\alpha_l+\frac{ \alpha_u-\alpha_l }{ N_\alpha-1 }i_\alpha \beta_l+\frac{ \beta_u-\beta_l }{ N_\beta-1 }i_\beta\right)^\mathrm{T}}$;
			\State 
			\begin{align*}
				\mathcal{V}'[i_\alpha][i_\beta]\leftarrow 
				\begin{cases}
					\displaystyle{ \min_{b\in\mathcal{B}} 
            \max_{a\in\mathcal{A}}\left[
                l \left( x ,a,b  \right)\Delta t + \widehat{ V}\left( x' \right)
			\right]   },  \ \ \ \ \text{for upper state-value function}\\
			\displaystyle{ \max_{a\in\mathcal{A}}\min_{b\in\mathcal{B}} 
            \left[
                l \left( x ,a,b  \right)\Delta t + \widehat{ V}\left( x' \right)
			\right]   },   \ \ \ \ \text{for lower state-value function}
				\end{cases}
			\end{align*}
          \EndFor
        \EndFor
        \State Copy $\mathcal{V}'$ to $\mathcal{V}$;
      \EndFor
      \State Return $\widehat{V}(.)$;
    \end{algorithmic}
\end{algorithm}

%定义纳什均衡算子\mathscr{N}
\subsection{Nash equilibrium state-value function}
To take the situations that the saddle point dose not exist into consideration,  
we introduce some notaions:
\begin{itemize}
	\item $ \mathscr{D}_{\mathcal{A}} $ and $ \mathscr{D}_{\mathcal{B}} $ represent the sets 
	of probability distributions on $\mathcal{A} $ and $\mathcal{B}$ respectively.
	\item $\hat{A}(.,.):[0,T] \to \mathscr{D}_{\mathcal{A}}$ and $\hat{B}(.,.):[0,T]\to \mathscr{D}_{\mathcal{B}}$ 
	represent the players'  mixed control inputs in the moving time frame. Specifically, $\hat{A}(t,.)$ and $\hat{B}(t,.)$ are the probability distributions 
	of the actions taken by player I and II at time $t$. $\hat{A}(t,a)$ is the probability of 
	player I choosing action $a$ at time $t$ and $\hat{B}(t,b)$ is the probability of 
	player II choosing action $b$ at time $t$.
	\item In a normal-form game, the game can be expressed as a game matrix. In this paper,
	the Nash equilibrium strategy pair under a given game matrix $\mathcal{G}$ is represented as 
	$\left(d_1^{\mathcal{G}*},d_2^{\mathcal{G}*}\right)$, where $d_1^{\mathcal{G}*}(a)$ is the probability of player I playing action $a$
	and $d_2^{\mathcal{G}*}(b)$
	is the probability of player II playing action $b$.
\end{itemize}

Then state-value function in the moving time frame 
under $\hat{A}(.,.)$ and $\hat{B}(.,.)$ 
can be expressed as
\begin{align}
	\begin{split}
	\hat{V}_T(x_0,\hat{A}(.,.),\hat{B}(.,.))=
	\mathbb{E}\bigg[ \int_0^T l(x(t),a(t),b(t))dt \bigg|
		a(t)\sim \hat{A}(t,.),
	b(t)\sim \hat{B}(t,.), x(0)=x_0
	\bigg]			
	\end{split}
\end{align}
Squently, the Nash equilibrium control input distribution pair $\left(\hat{A}^*(.,.),\hat{B}^*(.,.)\right)$ 
satisfy \cite{a25} \cite{a26}
\begin{align}
	\begin{split}
		\hat{A}^*(.,.)=\arg\max_{\hat{A}}\hat{V}_T(x_0,\hat{A}(.,.),\hat{B}^*(.,.))\\
		\hat{B}^*(.,.)=\arg\min_{\hat{B}}\hat{V}_T(x_0,\hat{A}^*(.,.),\hat{B}(.,.))
	\end{split}
\end{align}
% Given system state $x_0$, the actual control inputs are sampled from the initial distributions of $\hat{A}^*(.,.)$
% and $\hat{B}^*(.,.)$, that is 
% \begin{align}
% 	a\sim \hat{A}^*(0,.),b\sim \hat{B}^*(0,.)
% \end{align}
And the the Nash equilibrium state-value 
function in the moving time frame under Nash equilibrium control inputs is defined as 
\begin{align}
	\hat{V}_T^*(x_0)=\hat{V}_T(x_0,\hat{A}^*(.,.),\hat{B}^*(.,.))
\end{align}

Based on Equation (8),
we introduce the method to compute the state-value 
function in the moving time frame under Nash equilibrium control inputs.
In order to differentiate it from the upper and lower state-value functions, 
we introduce an operator $\mathscr{N}$ which act on a state-value function 
and return the Nash equilibrium state-value function. Mathematically, 
the Nash equilibrium state-value function is
\begin{align}
	\hat{V}_T^*(x_0)=\mathop{\mathscr{N}}_{\hat{A},\hat{B}}
	\hat{V}_T(x_0,\hat{A}(.,.),\hat{B}(.,.))
\end{align}
Similarly as Equation (5), for any $\tau\in (0,T) $, we have:
\begin{align}
	\begin{split}
	V_T^*(x_0)=&
	\mathop{\mathscr{N}}_{\hat{A},\hat{B}}
	\mathbb{E}\bigg\{\bigg[\int_0^\tau l(x(t),a(t),b(t))\ dt \bigg|a(t)\sim \hat{A}(t,.), 
	b(t)\sim \hat{A}(t,.),x(0)=x_0 \bigg] +\\
	&\mathop{\mathscr{N}}_{\hat{A}',\hat{B}'}
	\mathbb{E} \bigg[\int_\tau^T l(x(t),a'(t),b'(t))\ dt \bigg| a'(t)\sim \hat{A}'(t,.)
	b'(t)\sim \hat{B}'(t,.),x(\tau)\sim \mathcal{X}_0^\tau \left(x_0,\hat{A},\hat{B},.\right)   \bigg]
    \bigg\}\\
    =&\mathop{\mathscr{N}}_{\hat{A},\hat{B}}
	\mathbb{E}\bigg\{\bigg[\int_0^\tau l(x(t),a(t),b(t))\ dt \bigg|a(t)\sim \hat{A}(t,.), 
	b(t)\sim \hat{B}(t,.),x(0)=x_0 \bigg] +\\
	&\mathbb{E} \bigg[ V_{T-\tau}^*(x(\tau)) \bigg| x(\tau)\sim \mathcal{X}_0^\tau \left(x_0,\hat{A},\hat{B},.\right) \bigg] 
	\bigg\}
	\end{split}
\end{align}
where $\mathcal{X}_0^\tau \left(x_0,\hat{A},\hat{B},.\right)$ is a probability distribution,
and $\mathcal{X}_0^\tau \left(x_0,\mathcal{S}_0^\tau,\mathcal{T}_0^\tau,x_1\right)$ is the probability of 
system state transition from $x_0$ to $x_1$ under control inputs $\hat{A}(.,.)$ and $\hat{B}(.,.)$ 
over time interval $[0,\tau]$. %See Fig. 1 for the visual explanation.

Similarly as Equation (7), with a small enough $\Delta t$, the follows hold:
\begin{align}
	\begin{split}
		&\hat{V}_{\Delta t}^*(x_0)=\mathop{\mathscr{N}}_{ d_1\in \mathscr{D}_{\mathcal{A}} ,d_2\in \mathscr{D}_{\mathcal{B}} }
		\mathbb{E}\bigg[l\left( x_0,a,b  \right)\Delta t \bigg| a\sim d_1,
		b\sim d_2 \bigg] \\
		&\hat{V}_{k\Delta t}^*(x_0)=\mathop{\mathscr{N}}_{d_1\in \mathscr{D}_{\mathcal{A}} ,d_2\in \mathscr{D}_{\mathcal{B}} }
		\mathbb{E}\bigg[l\left(x_0,a,b  \right)\Delta t 
		+\hat{V}_{(k-1)\Delta t}^*\left( F(x_0,a,b)\right) \bigg|a\sim d_1, b\sim d_2 \bigg]
	\end{split}
\end{align}
See Fig.1 for the visual descriptions of Equation (13).

\begin{figure}[H]
	\centering
	\subfigure[\footnotesize{Entire game tree}]{\includegraphics[height=5cm]{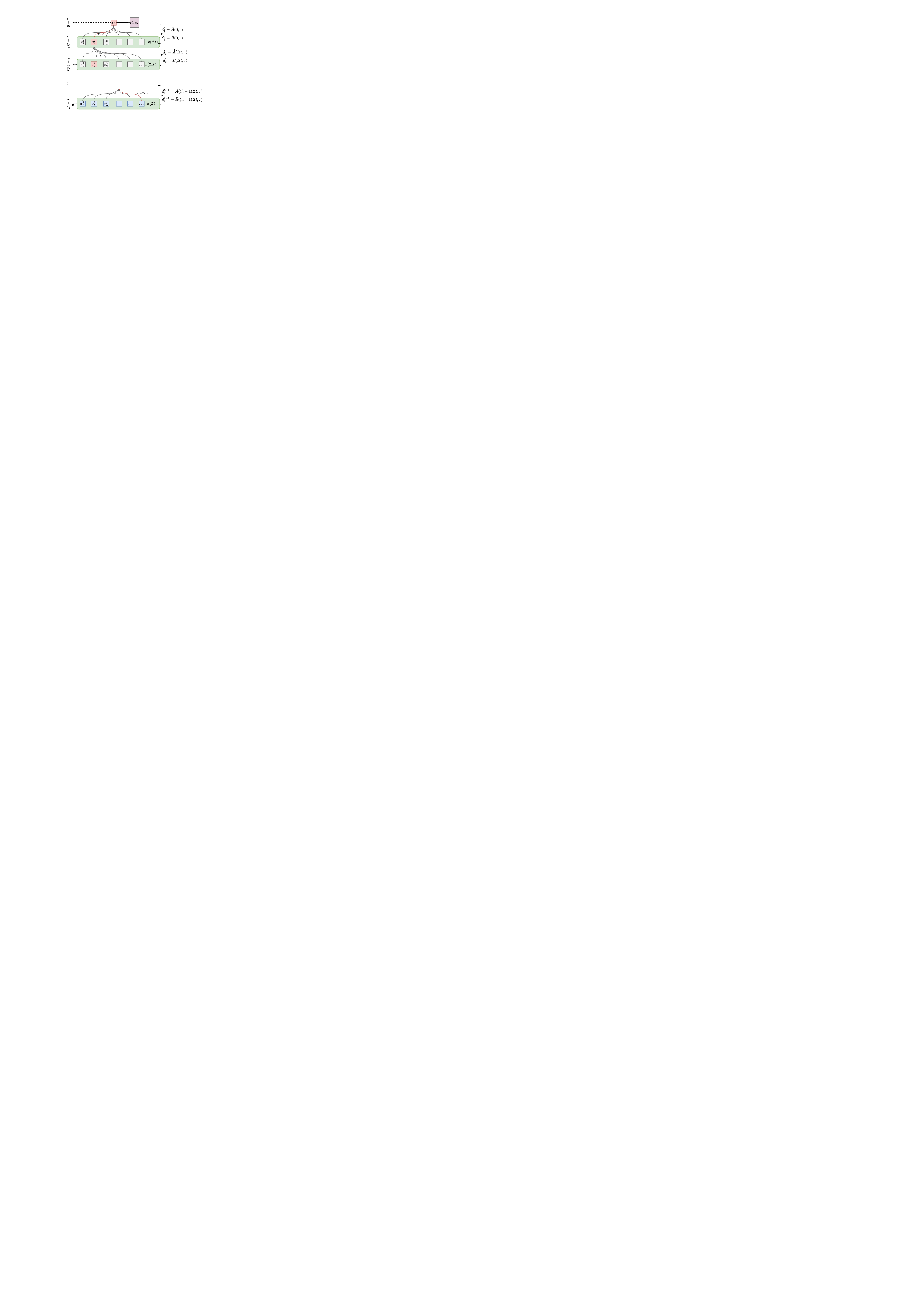} } 
	\subfigure[\footnotesize{Reduced game tree}]{\includegraphics[height=5.3cm]{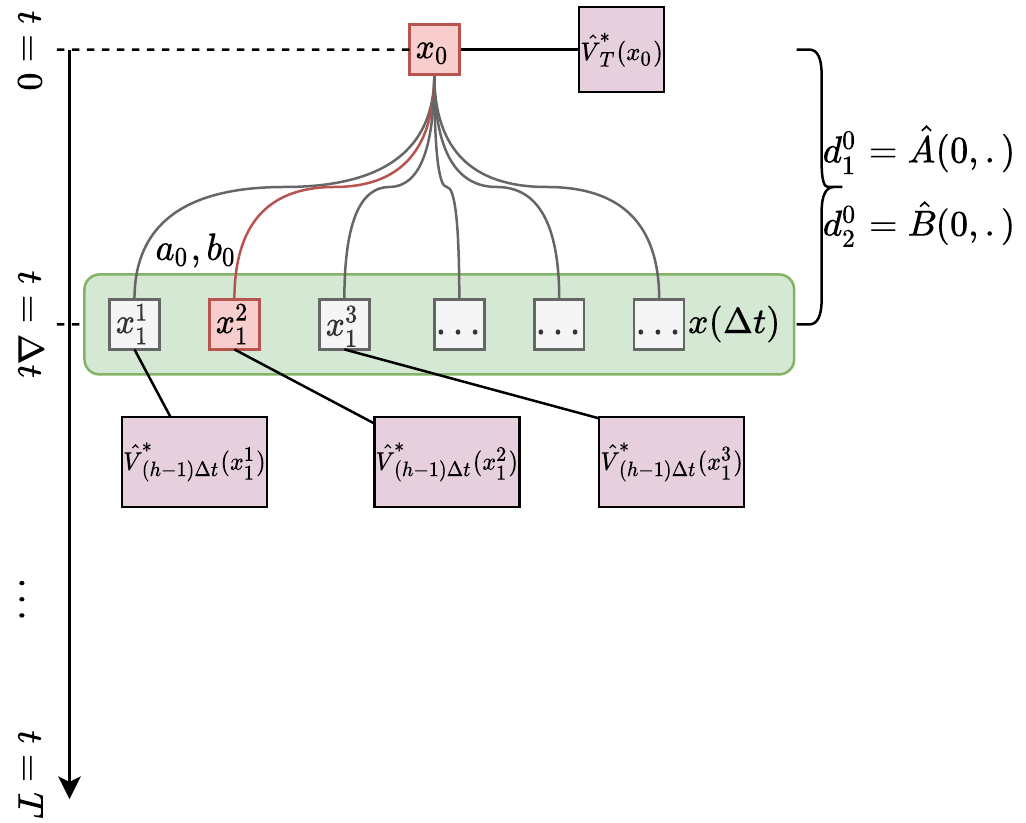} } \ \ \ \ \ \ \\
	\subfigure[\footnotesize{Recursive form}]{\includegraphics[height=4.5cm]{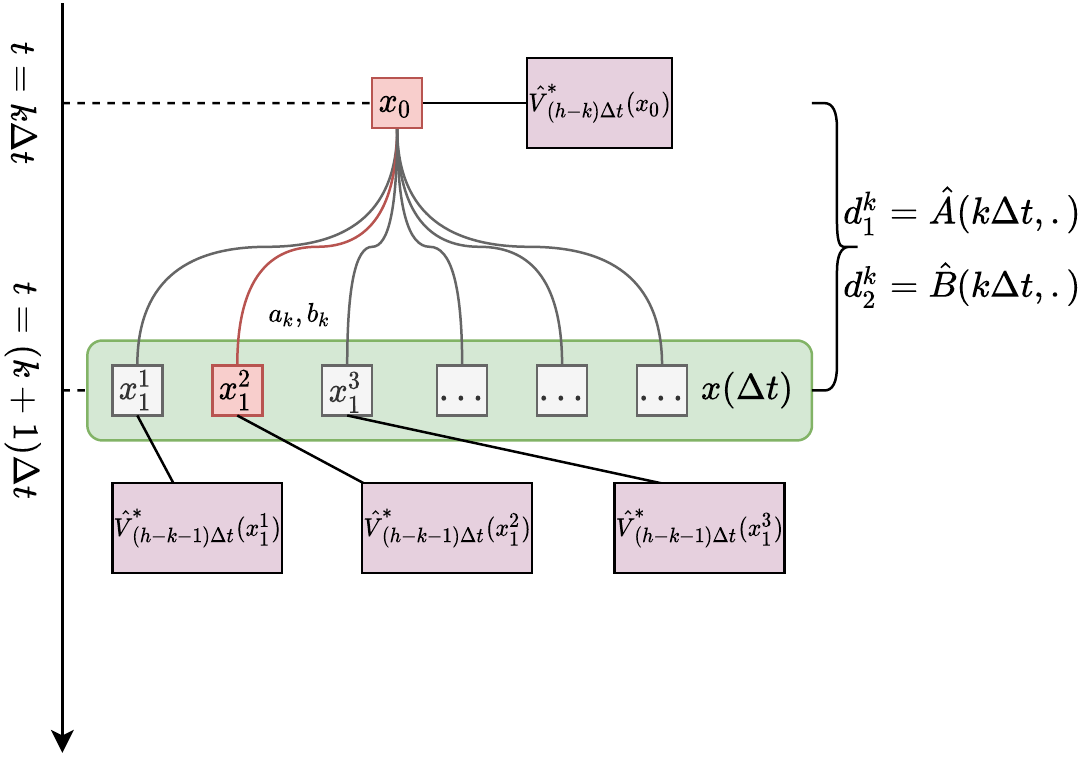} } \ \ \
	\caption{(a) To compute $\hat{V}^*_T(x_0)$, we need to traverse the entire game tree, then 
	$\hat{V}^*_T(x_0)= \mathscr{N}_{d_1^0,...,d_1^{h-1},
	d_2^0,...,d_2^{h-1} }\mathbb{E} \big\{ [l(x_0,a_0,b_0) + 
	l(x(\Delta t),a_1,b_1)+...$ $+ l(x((h-1)\Delta t),a_{h-1},b_{h-1}) ]\Delta t \big| a_0\sim d_1^0,...,
	a_{h-1}\sim d_1^{h-1}, b_0\sim d_2^0,...,
	b_{h-1}\sim d_2^{h-1}  \big\} $. 
	(b) Given the state-value function $\hat{V}^*_{(h-1)\Delta t}(.)$, 
	$\hat{V}^*_T(x_0)=\mathscr{N}_{d_1^0,
	d_2^0 }\mathbb{E} [ l(x_0,a_0,b_0)\Delta t+ \hat{V}^*_{(h-1)\Delta t}(x(\Delta t)) | a_0\sim d_1^0, 
	b_0\sim d_2^0 ]$.
	(c) For any $k\in\{1,...,h\}$, we can obtain $\hat{V}^*_{(h-k)\Delta t}(.)$ via recursion: 
	$\hat{V}^*_{(h-k)\Delta t}(x_0)=\mathscr{N}_{d_1^k,
	d_2^k }\mathbb{E} [ l(x_0,a_k,b_k)\Delta t+ \hat{V}^*_{(h-k-1)\Delta t}(x(\Delta t)) | a_k\sim d_1^k, 
	b_k\sim d_2^k ]$.}
\end{figure}
% The goal is to compute $ \hat{V}_{h\Delta t}^*(.) $. 
According to Equation (13), for any $m=1,2,...$, $\hat{V}_{m\Delta t}^*(.)$ can be obtained by $m$-step recursion.
There are infinite elements in set $\mathcal{A}$ and $\mathcal{B}$, thus,
both players have infinite actions to be taken. However, 
traversing the entire $\mathcal{A}$ and $\mathcal{B}$ is impossible.
Fortunately, many of those actions are very similar,
we can just consider two sets of finite elements that distribute evenly in 
$\mathcal{A}$ and $\mathcal{B}$ instead of the entire ones.
This is referred to as action abstraction \cite{a21} \cite{a22}.
We represent the action abstractions of $\mathcal{A}$ and $\mathcal{B}$ as $\bar{\mathcal{A}}$ and $\bar{\mathcal{B}}$
respectively and denote the $i$th element in $\bar{\mathcal{A}}$ and $\bar{\mathcal{B}}$ as 
$\bar{\mathcal{A}}[i]$ and $\bar{\mathcal{B}}[i]$ respectively.

Given the state-value function $\hat{V}_{(k-1)\Delta t}^*(.)$ and the the action abstractions $\bar{\mathcal{A}}$ and $\bar{\mathcal{B}}$, 
the second equation in (13) can be translated into a \emph{normal-form game}:  
Construct a $N_a\times N_b$ game matrix $\hat{\mathcal{G}}_k$
($N_a$ and $N_b$ are the numbers of elements in $\bar{\mathcal{A}}$ and $\bar{\mathcal{B}}$), 
where each dimension has rows/columns corresponding to a single player's actions \cite{a24}.
By convention, the row player is player I and the column player
is player II, and fill the entry in row $i$ column $j$ with
\begin{align}
	\hat{\mathcal{G}}_k[i][j]=l\left( x_0,\bar{\mathcal{A}}[i],\bar{\mathcal{B}}[j]  \right)\Delta t 
	+\hat{V}_{(k-1)\Delta t}^*\left( F(x_0,\bar{\mathcal{A}}[i],\bar{\mathcal{B}}[j])\right)
\end{align} 
A part of the game matrix $\hat{\mathcal{G}}_k$ is shown in Fig. 2.
\begin{figure}[H]
	\centering
	\includegraphics[width=6cm]{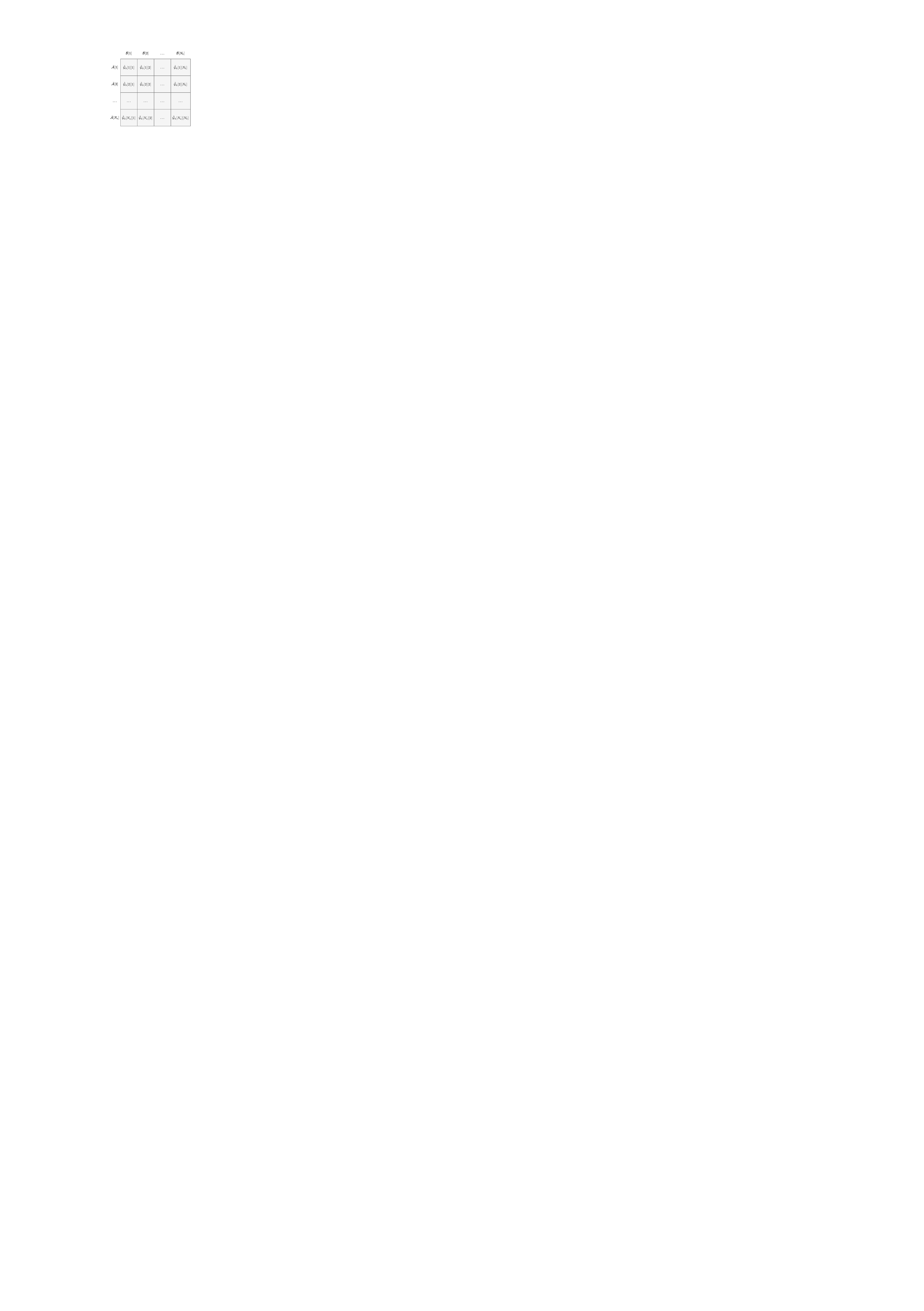}
	\caption{The game matrix $\hat{\mathcal{G}}_k$.}
\end{figure}
Then the regret matching can be adopted to compute the Nash equilibrium
strategies, $d_{1}^{\hat{\mathcal{G}}_k*} \in \mathscr{D}_{\bar{\mathcal{A}}}$ and 
$d_{2}^{\hat{\mathcal{G}}_k*} \in \mathscr{D}_{\bar{\mathcal{B}}}$, for both players.
% We express the Nash equilibrium strategies under the given game matrix $\hat{\mathcal{G}}$ as 
% a pair of probability distributions 
% $d_{1}^{\hat{\mathcal{G}}*} \in \mathscr{D}_{\bar{\mathcal{A}}}$ and 
% $d_{2}^{\hat{\mathcal{G}}*} \in \mathscr{D}_{\bar{\mathcal{B}}}$, 
% where $d_{1}^{\hat{\mathcal{G}}*}(a) $ is the probability of player I playing action $a$, 
% and $d_{2}^{\hat{\mathcal{G}}*}(b)$ is the probability of player II playing action $b$. 
% ---A probability distribution on $\bar{\mathcal{A}}$ or $\bar{\mathcal{B}}$.
% Let $\bar{\mathcal{S}}^*_\mathcal{G}(.)$ and $\bar{\mathcal{T}}^*_\mathcal{G}(.)$ denote 
% the Nash equilibrium strategies for player I and player II, $\bar{\mathcal{S}}^*_\mathcal{G}(a)$ 
% represents the probability of player I taking action $a$ and 
% $\bar{\mathcal{T}}^*_\mathcal{G}(b)$ 
% represents the probability of player II taking action $b$.
Then, to compute $\hat{V}_{k\Delta t}^*(x_0)$, sum over each action pair,
the product of each player's probability of playing their action in the action pair, times 
the value in the corresponding entry:
\begin{align}
	\hat{V}_{k\Delta t}^*(x_0)=\sum_{i=1}^{N_a}\sum_{j=1}^{N_b} 
	d_{1}^{\hat{\mathcal{G}}_k*}(\bar{\mathcal{A}}[i]) d_{2}^{\hat{\mathcal{G}}_k*}(\bar{\mathcal{B}}[j])
	\hat{\mathcal{G}}_k[i][j]
\end{align}

Squently, making a little change on Algorithm 1, we can obtain an approximation 
of the Nash equilibrium state-value function $\hat{V}_{m\Delta t}^*(.)$ for $m=1,2,...$, see Algorithm 2
(with input $m=h$, we can obtain $\hat{V}_{T}^*(.)$).
\begin{algorithm}[h]
    \caption{Computation of Nash equilibrium state-value function}
	\begin{algorithmic}[1]
	  \State \textbf{Input:} Recursion number $m$;
	  \State Select a proper rectangular computational domain $\Omega=[\alpha_l,\alpha_u]\times [\beta_l,\beta_u]$; 
      \State Discretize $\Omega$ into a $N_\alpha\times N_\beta $ Cartesian grid structure;
	  \State Let $\mathcal{V}$ and $\mathcal{V}'$ be two $N_\alpha\times N_\beta$ arrays 
	  and initialize them to $\mathbf{0}$;
    %   and let $\mathcal{V}[i_\alpha][i_\beta]$ denote the state-value of state 
	%   $\displaystyle{x_{i_\alpha,i_\beta}=\left(\alpha_l+\frac{ \alpha_u-\alpha_l }{ N_\alpha-1 }i_\alpha,
	%   \beta_l+\frac{ \beta_u-\beta_l }{ N_\beta-1 }i_\beta\right)^\mathrm{T}}$; 
      \For{$k \leftarrow 1,...,m$}
        \For{$i_\alpha\leftarrow 0,...,N_\alpha-1 $}
		  \For{$i_\beta\leftarrow 0,...,N_\beta-1 $}
		  	\State $x\leftarrow \displaystyle{\left(\alpha_l+\frac{ \alpha_u-\alpha_l }{ N_\alpha-1 }i_\alpha \beta_l+\frac{ \beta_u-\beta_l }{ N_\beta-1 }i_\beta\right)^\mathrm{T}}$;
			\State Construct a $N_a\times N_b$ game matrix $\hat{\mathcal{G}}_k$;
			\For{$i \leftarrow 1,...,N_a $}
				\For{$j \leftarrow 1,...,N_b $}
					\State $a\leftarrow \bar{\mathcal{A}}[i]$;
					\State $b\leftarrow \bar{\mathcal{B}}[j]$;
					\State $x'=F(x,a,b)$;
					\State $\hat{\mathcal{G}}_k[i][j]\leftarrow   
					l\left( x,a,b  \right)
					\Delta t 
					+\widehat{V}\left( x'\right)$;
				\EndFor
			\EndFor
			\State Use regret matching to compute $d_{1}^{\hat{\mathcal{G}}_k*}(.)$ and $d_{2}^{\hat{\mathcal{G}}_k*}(.)$;
			\State \begin{align*}
				\mathcal{V}'[i_\alpha][i_\beta]\leftarrow \displaystyle{ \sum_{i=1}^{N_a}\sum_{j=1}^{N_b} 
			d_{1}^{\hat{\mathcal{G}}_k*}(\bar{\mathcal{A}}[i]) d_{2}^{\hat{\mathcal{G}}_k*}(\bar{\mathcal{B}}[j])
			\hat{\mathcal{G}}_k[i][j]};
			\end{align*}
			% $\mathcal{V}'[i_\alpha][i_\beta]\leftarrow \\ \displaystyle{ \sum_{i=1}^{N_a}\sum_{j=1}^{N_b} 
			% d_{1}^{\hat{\mathcal{G}}_k*}(\bar{\mathcal{A}}[i]) d_{2}^{\hat{\mathcal{G}}_k*}(\bar{\mathcal{B}}[j])
			% \hat{\mathcal{G}}_k[i][j]}$;
          \EndFor
        \EndFor
        \State Copy $\mathcal{V}'$ to $\mathcal{V}$;
      \EndFor
      \State Return $\widehat{V}(.)$;
    \end{algorithmic}
\end{algorithm}

\subsection{Control policy}
\par At time $t$, the actual optimal control inputs should be sampled from 
the initial distributions of $\hat{A}^*(.,.)$ and $\hat{B}^*(.,.)$, 
that is 
\begin{align}
	\begin{split}
		a(t)\sim \hat{A}^*(0,.),b(t)\sim \hat{B}^*(0,.)
	\end{split}
\end{align}
where 
\begin{align}
	\begin{split}
		(\hat{A}^*(.,.),\hat{B}^*(.,.))=\arg \mathop{\mathscr{N}}_{\hat{A},\hat{B}}
		\hat{V}_T(x(t),\hat{A}(.,.),\hat{B}(.,.))
	\end{split}
\end{align}
According to Equation (13), with a small enough $\Delta t$, 
\begin{align}
	(\hat{A}^*(0,.),\hat{B}^*(0,.))=\arg\mathop{\mathscr{N}}_{d_1\in \mathscr{D}_{\mathcal{A}} ,d_2\in \mathscr{D}_{\mathcal{B}} }
	\mathbb{E}\bigg[l\left(x(t),a,b  \right)\Delta t 
	+\hat{V}_{(h-1)\Delta t}^*\left( F(x(t),a,b)\right) \bigg|a\sim d_1, b\sim d_2 \bigg]
\end{align}
holds.
Therefore, 
given the state-value function $\hat{V}^*_{m\Delta t}(.) $ generated by Algorithm 2 with $m=h-1$, 
$\hat{A}^*(0,.)=d_1^{\mathcal{G}*} $ and $\hat{B}^*(0,.)=d_2^{\mathcal{G}*}$,
% the actual optimal control inputs of both players at time $t$ can be sampled from 
% $d_1^{\mathcal{G}*} $ and $d_2^{\mathcal{G}*} $, 
where $\mathcal{G}$ is a game matrix, and 
\begin{align}
	\mathcal{G}[i][j]=l\left( x(t),\bar{\mathcal{A}}[i],\bar{\mathcal{B}}[j]  \right)\Delta t 
	+\hat{V}^*_{(h-1)\Delta t}\left( F(x(t),\bar{\mathcal{A}}[i],\bar{\mathcal{B}}[j])\right)
\end{align}
% The Nash equilibrium strategy pair $(d_1^{\mathcal{G}^T*} ,d_2^{\mathcal{G}^T*})$ under $\mathcal{G}^T$ can be easily 
% solved via regret matching.
% $\bar{\mathcal{S}}^*_{\mathcal{G}^T}(.)$ and $\bar{\mathcal{T}}^*_{\mathcal{G}^T}(.)$ can be obtained via 
% regret matching. 

In the question of on-line control, the state-value function $\hat{V}_{(h-1)\Delta t}(.)$
can be held in memory. Then construct a game matrix $\mathcal{G}$ and fill it using Equation (19). 
Finally, the Nash equilibrium strategy pair $(d_1^{\mathcal{G}*} ,d_2^{\mathcal{G}*})$ can be obtained via 
regret matching. Under the Nash equilibrium, 
no player can increase its own expected payoff (the performance index in the moving time frame for player I and the minus of it for player II)
by changing only their own strategy, therefore, each player just needs to choose 
its own action according to its onw Nash equilibrium strategy and does not need to focus on the action taken by its opponent.
Take player I as an example, its control block diagram is shown in Fig. 3.
\begin{figure}[H]
	\centering
	\includegraphics[width=7cm]{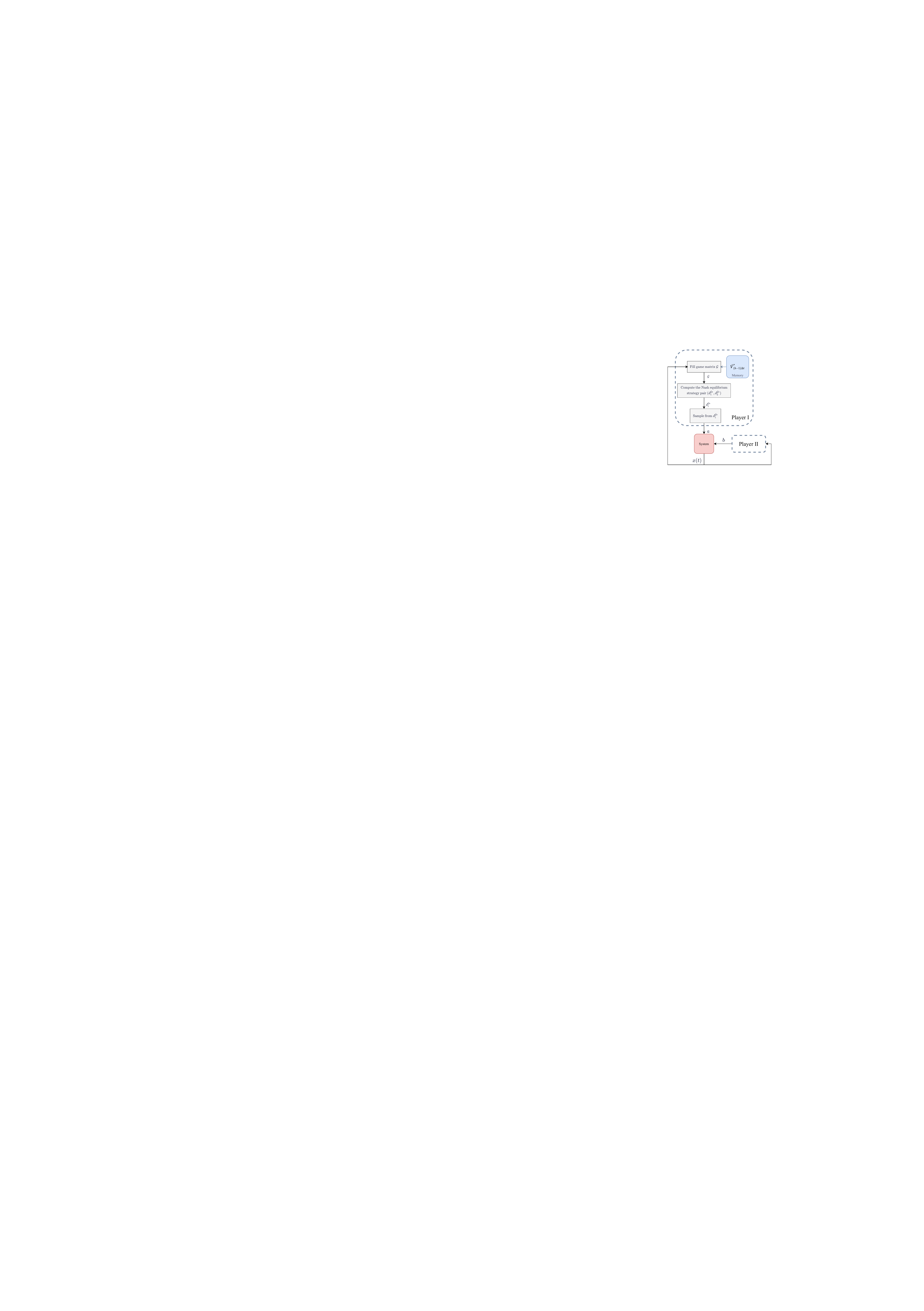}
	\caption{Control block diagram of player I.}
\end{figure}
% With keeping $\hat{V}_T^*(.) $ in memory, the control policies of both 
% players can be defined as two mappings: $\Pi^{\hat{V}_T^*}_1(.,.):\mathbb{R}^n\times \bar{\mathcal{A}} \to \mathbb{R}$,
% $\Pi^{\hat{V}_T^*}_2(.,.):\mathbb{R}^n\times \bar{\mathcal{B}} \to \mathbb{R}$. 
% And $\Pi^{\hat{V}_T^*}_1(x_0,a)$ is the probability of player I choosing action $a$ at state $x_0$, 
% $\Pi^{\hat{V}_T^*}_2(x_0,b)$ is the probability of player II choosing action $b$ at state $x_0$.

\section{Numerical Examples}
\par The dynamics of the benchmark nonlinear plant can
be expressed by
\begin{align}
	\dot{x}=\left[
		\begin{array}{c}
			\dot{\alpha} \\
			\dot{\beta}
		\end{array}
	\right]=\left[
		\begin{array}{c}
			-(\alpha+\beta+a)^3 \\
			-(\beta-\alpha+b)^3
		\end{array}
	\right]=f(x,a,b)
\end{align}
where $a\in \mathcal{A}=[-1,1]$, $b\in \mathcal{B}=[-1,1]$.

\subsection{Example 1}
The running payoff function is expressed as:
\begin{align}
	l(x,a,b)=1+\alpha^2-\beta^2-a^2+b^2
\end{align}
The time step size is set as $\Delta t=0.02$.
The computational domain is set as $\Omega=[-1,1]\times [-1,1]$. 
We discretize $\Omega$ into a $257\times 257$ Cartesian grid structure.
And the action abstractions are 
\begin{align}
	\bar{\mathcal{A}}=\bar{\mathcal{B}}=\left\{-1+0.05i|i=0,...,40   \right\}
\end{align}
\subsubsection{Computation of state-value functions}
Given a $T$, the upper and lower state-value functions $V_T^u(.)$ and $V_T^l(.)$ 
can be approximated via Algorithm 1. And the Nash equilibrium state-value
function $V_T^*(.)$ can be approximated via Algorithm 2. 
Fig. 4 shows $V_T^u(.)$, $V_T^l(.)$ and $V_T^*(.)$ under $T=1$ (in Algorithm 1 and Algorithm 2, input $m=50$). 
It can be seen that $V_1^u(.)=V_1^l(.)$, that 
means the saddle point exists. 
Therefore, as expected, $V_1^u(.)=V_1^l(.)=V_1^*(.)$. 
We show the optimal actual control inputs and game matrix at states $(0.5,0.5)^\mathrm{T}$ and $(0.3,0.8)^\mathrm{T}$ 
with $V_{1}^*(.)$ kept in memory, %as a graphical display, 
see Fig. 5. As expected, both players choose a single action with probability 1.

\begin{figure}[H]
	\centering
	% \subfigure[\footnotesize{$V^u_{0.5}(.) $}]{\includegraphics[height=6cm]{sv5.png} }   \\
	% \subfigure[\footnotesize{$V^l_{0.5}(.) $}]{\includegraphics[height=6cm]{sv5.png} }  \\
	% \subfigure[\footnotesize{$V^*_{0.5}(.) $}]{\includegraphics[height=6cm]{sv5.png} }  
	\subfigure[\footnotesize{$V^u_1(.) $}]{\includegraphics[height=6cm]{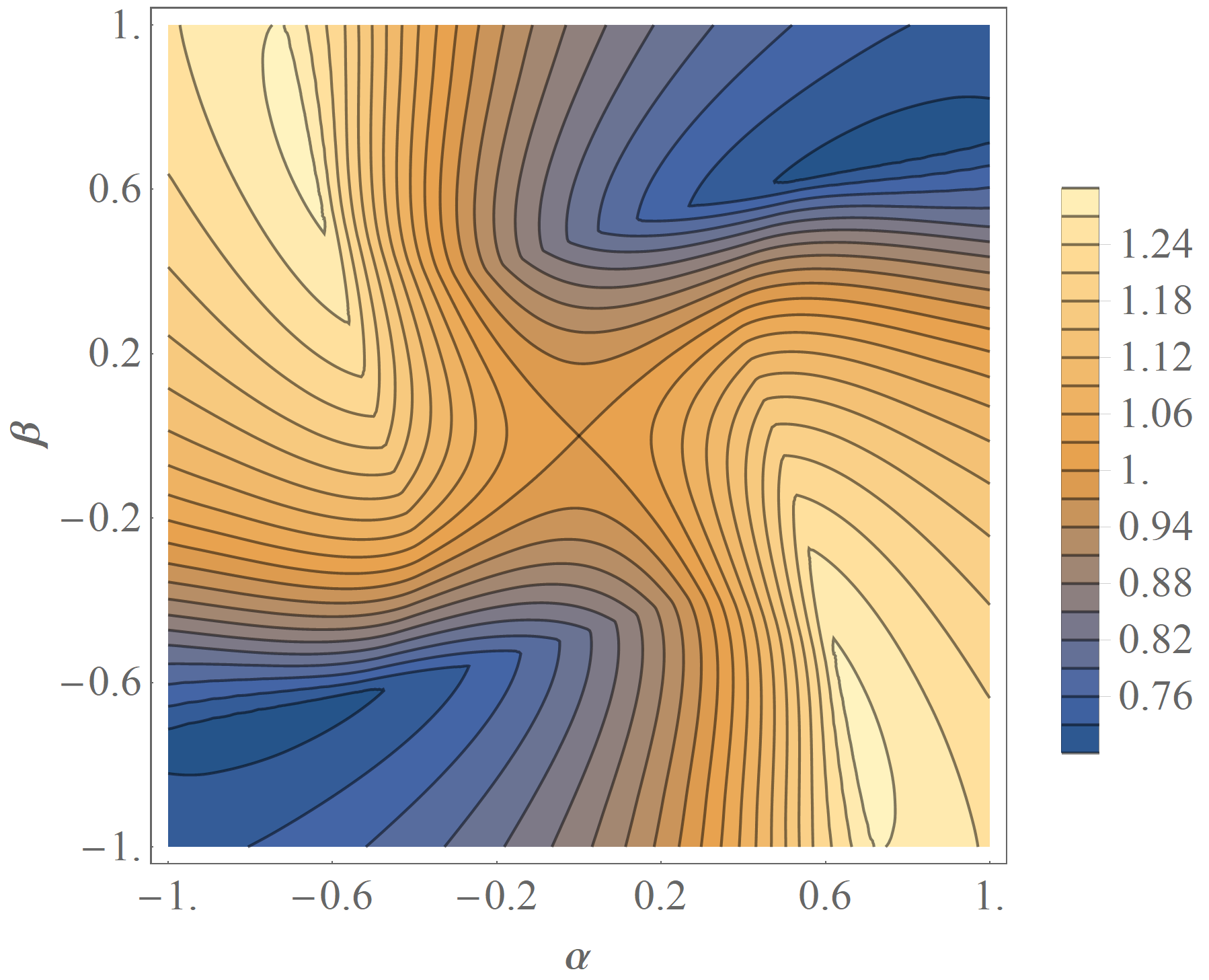} }   
	\subfigure[\footnotesize{$V^l_1(.) $}]{\includegraphics[height=6cm]{sv10.png} }  \\
	\subfigure[\footnotesize{$V^*_1(.) $}]{\includegraphics[height=6cm]{sv10.png} }  
	% \subfloat[\footnotesize{State-value function}]{\includegraphics[height=4.5cm]{sv15.png} }   
	% \subfloat[\footnotesize{State-value function}]{\includegraphics[height=4.5cm]{sv15.png} }  
	% \subfloat[\footnotesize{State-value function}]{\includegraphics[height=4.5cm]{sv15.png} }  \\
	% \subfigure[\footnotesize{$V^u_2(.) $}]{\includegraphics[height=4.5cm]{sv20.png} }   
	% \subfigure[\footnotesize{$V^l_2(.) $}]{\includegraphics[height=4.5cm]{sv20.png} }  
	% \subfigure[\footnotesize{$V^*_2(.) $}]{\includegraphics[height=4.5cm]{sv20.png} }  
	\caption{The upper, lower and Nash equilibrium state-value functions under $T=1$.}
\end{figure}

\begin{figure}[H]
	\centering
	% \subfloat[\footnotesize{State-value function}]{\includegraphics[height=4.5cm]{sv5.png} }   
	% \subfloat[\footnotesize{State-value function}]{\includegraphics[height=4.5cm]{sv5.png} }  
	% \subfloat[\footnotesize{State-value function}]{\includegraphics[height=4.5cm]{sv5.png} }  \\
	\subfigure[\footnotesize{Optimal actual control inputs and game matrix at state $(0.5,0.5)^\mathrm{T}$}]{\includegraphics[height=6.9cm]{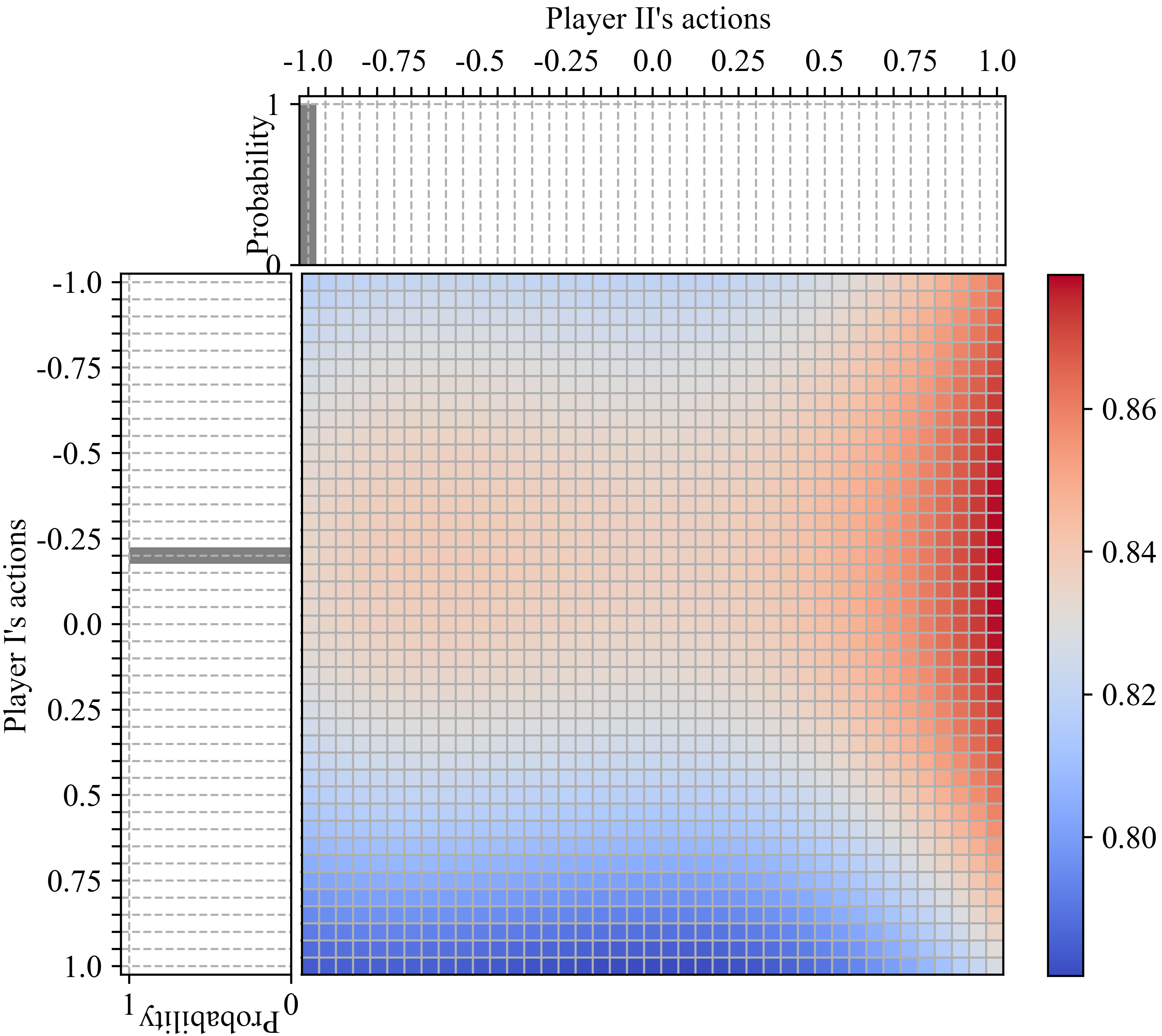} } 
	\subfigure[\footnotesize{Optimal actual control inputs and game matrix at state $(0.3,0.8)^\mathrm{T}$}]{\includegraphics[height=6.9cm]{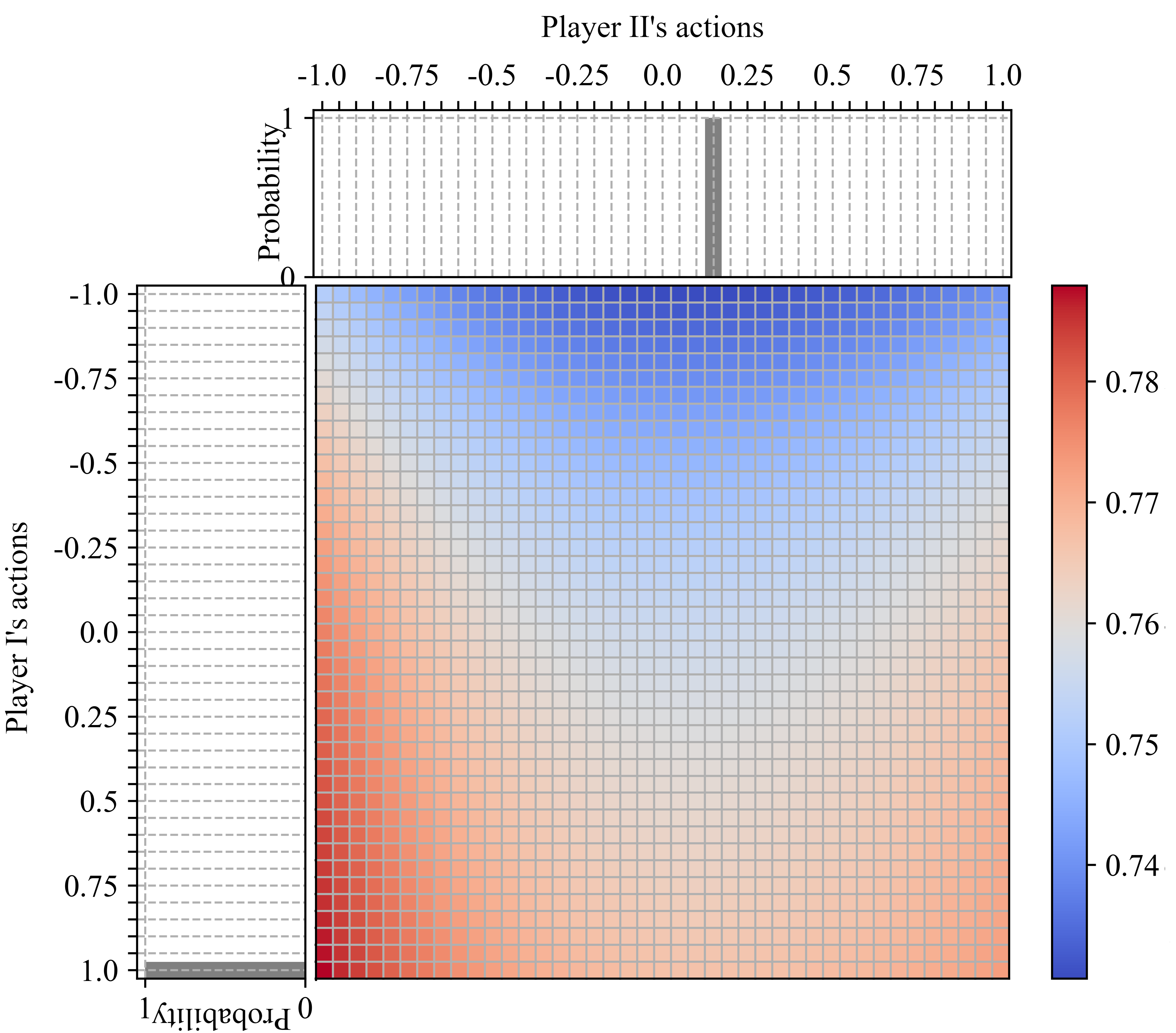} }  
	\caption{(a) At state $(0.5,0.5)^\mathrm{T}$, player I chooses the single action $a=0.2$ with probability 1, 
	player II chooses the single action $b=0$ with probability 1.
	(b) At state $(0.3,0.8)^\mathrm{T}$, player I chooses the single action $a=1$ with probability 1, 
	player II chooses the single action $b=0.15$ with probability 1. The values in different entries of the game matrix 
	are expressed as different colors.}
\end{figure}

\subsubsection{Battle between different control policies}
To demonstrate the effectiveness of our method, 
some battles between different control policies are simulated. 
These optional control policies include our method and the control policies 
with the existence hypothesis of saddle points. 
The later can be categorized into two types: 
\begin{itemize}
	\item The Min-max type: This type aims to find the optimal control input corresponding to 
	the upper state-value function.
	\item The Max-min type: This type aims to find the optimal control input corresponding to 
	the lower state-value function.
\end{itemize}
The rolling optimization with predictive period $T=1$ are adopted in all the control policies. 
When a player take our method, the state-value function $\hat{V}^*_{0.98}(.)$ computed by
Algorithm 2 (input $m=49$) is kept in its memory. For each control policy pair,
we take 100 simulations with random initial states and the average value of
these simulations' cumulative payoff is calculated.
The duration of these simulations are set as 10.
The process of a battle between two control policies is shown in Fig. 6. 
\begin{figure}[H]
	\centering
	\includegraphics[width=6cm]{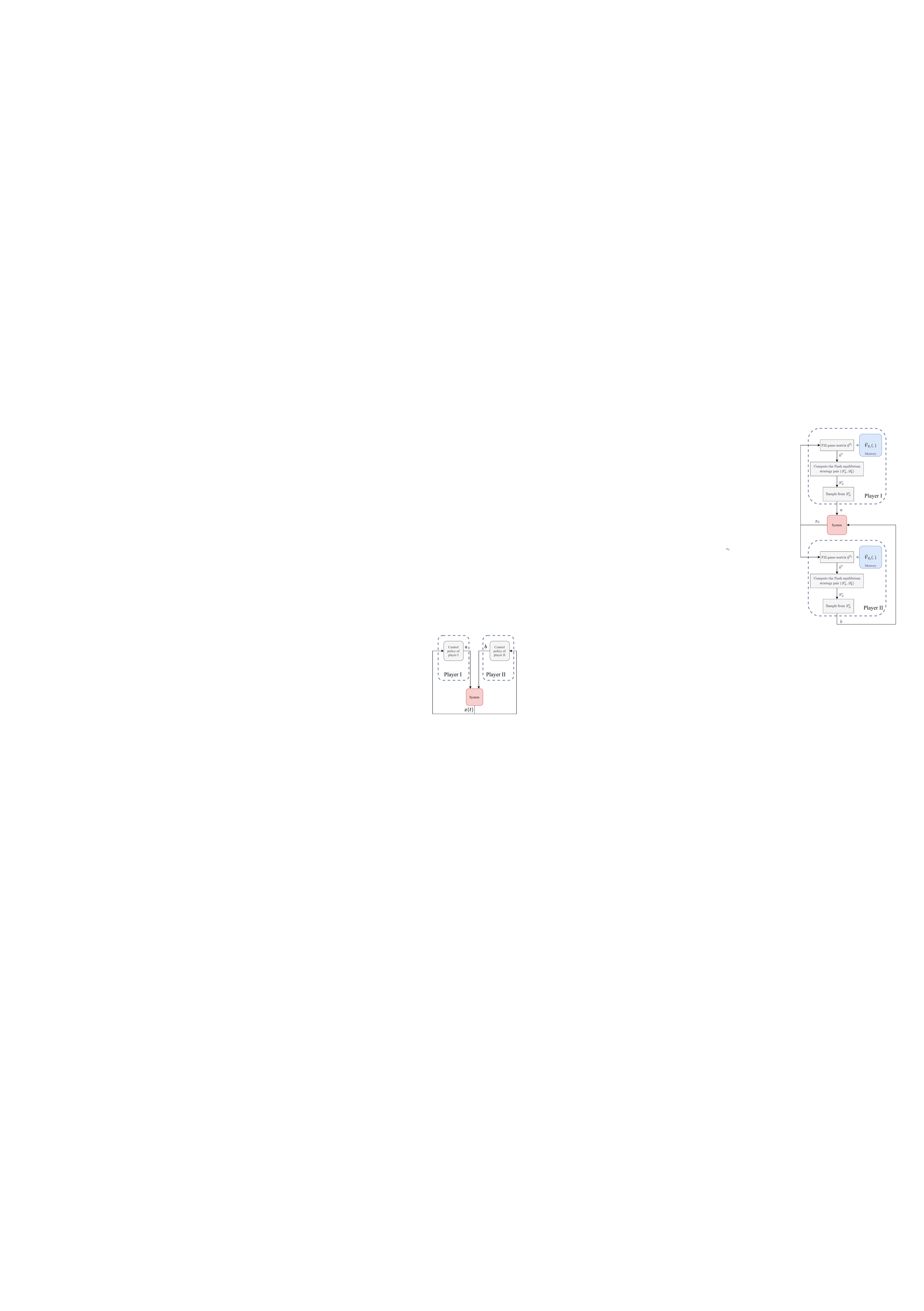}
	\caption{Block diagram of a battle.}
\end{figure}
% According to Equation (14), the game matrix $\mathcal{G}^T$ relies on $T$. 
% Therefore, different $T$ can generate different strategy pair 
% $\bar{\mathcal{S}}^*_{\mathcal{G}^T}(.)$ and $\bar{\mathcal{T}}^*_{\mathcal{G}^T}(.)$. 
% $\bar{\mathcal{S}}^*_{\mathcal{G}^T}(.)$ and $\bar{\mathcal{T}}^*_{\mathcal{G}^T}(.)$ 
% are the Nash equilibrium strategies for current actions with the consideration 
% of the performance index over a period of future time $T$. Thus, they rely on 
% $T$. 
The battle results are shown in Table 1. 
It can be seen that the values in different entries of Table 1 have no significant difference, 
this happened due to the existence of the saddle point, the control inputs generated by these three
control policies are same. 
\begin{table}[H]
	\centering
	\caption{Battle results of Example 1.}
	\begin{tabular}{cc|c|c|c|}
	\cline{3-5}
											  &   & \multicolumn{3}{c|}{Player II's control policy} \\ \cline{3-5} 
											  &   & Min-max type    & Max-min type      & Our method      \\ \hline
	\multicolumn{1}{|l|}{\multirow{3}{*}{\rotatebox{90}{Player I's control policy}    }} & \rotatebox{90}{Min-max type} &    10.03673    &   10.03351     &    10.03608   \\ \cline{2-5} 
	\multicolumn{1}{|l|}{}                    & \rotatebox{90}{Max-min type} &    10.0347    &    10.0284    &    10.0306   \\ \cline{2-5} 
	\multicolumn{1}{|l|}{}                    & \rotatebox{90}{Our method} &    10.0293    &    10.0325    &    10.0289  \\ \hline
	\end{tabular}
\end{table}

\subsection{Example 2}
In this example, the running payoff function is changed as
\begin{align}
	l(x,a,b)=1+10\sin (4ab)+\alpha^2-\beta^2
\end{align}
All other settings are the same
as the ones in Example 1. 

\subsubsection{Computation of state-value functions}
\par The upper and lower state-value functions under predictive period $T=1$s 
are shown in Fig. 7(a) and (b). 
It can be seen that 
$V_1^u(.)\neq V_1^l(.)$, that 
means the saddle point does not exist, 
we can only use Algorithm 2 to compute the Nash equilibrium state-value function, 
see Fig. 7(c). 
In order to visually display the mixed optimal actual control inputs, 
we show the optimal actual control inputs and game matrix
at states $(0.2,-0.2)^\mathrm{T}$ and $(0.3,0.35)^\mathrm{T}$ with $V_{1}^*(.)$ kept in memory, 
see Fig. 8. Since the saddle point dose not exist, both players 
have many actions that are played with positive probability.

\begin{figure}[H]
	\centering
	% \subfigure[\footnotesize{$V^u_1(.) $}]{\includegraphics[height=6cm]{svu5.png} }   \\
	% \subfigure[\footnotesize{$V^l_1(.) $}]{\includegraphics[height=6cm]{svl5.png} }  \\
	% \subfigure[\footnotesize{$V^*_1(.) $}]{\includegraphics[height=6cm]{svn5.png} }  
	\subfigure[\footnotesize{$V^u_1(.) $}]{\includegraphics[height=6cm]{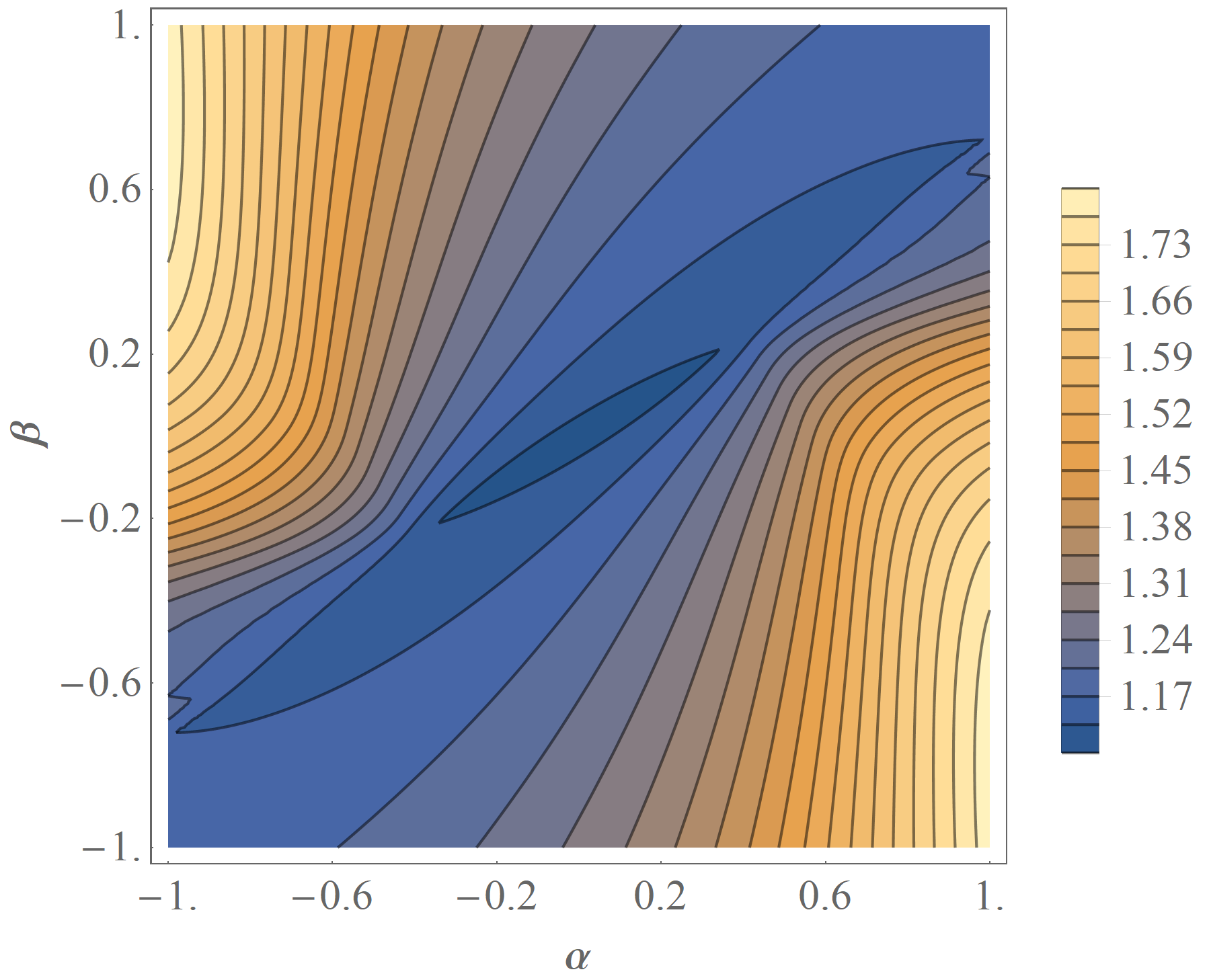} }   
	\subfigure[\footnotesize{$V^l_1(.) $}]{\includegraphics[height=6cm]{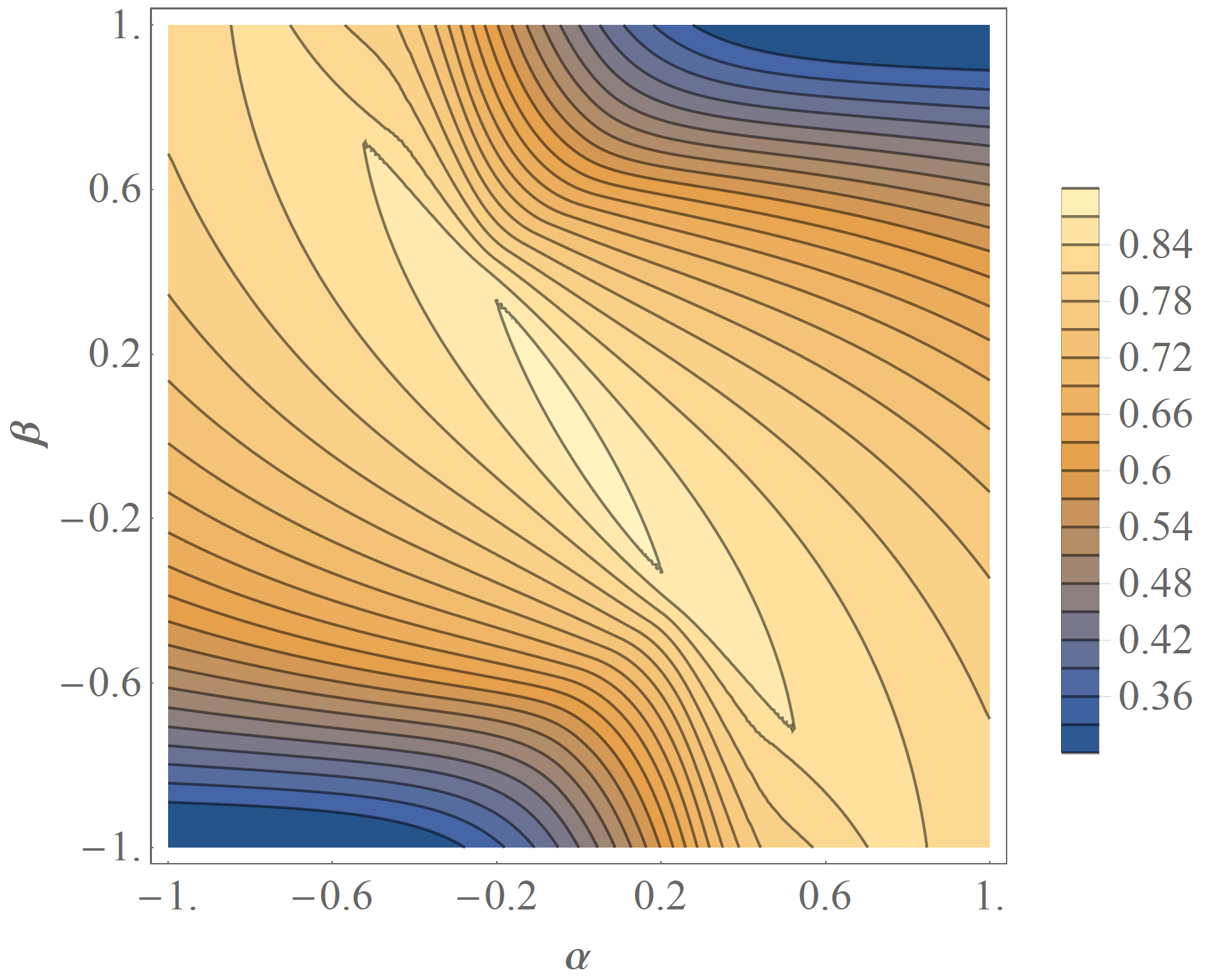} }  \\
	\subfigure[\footnotesize{$V^*_1(.) $}]{\includegraphics[height=6cm]{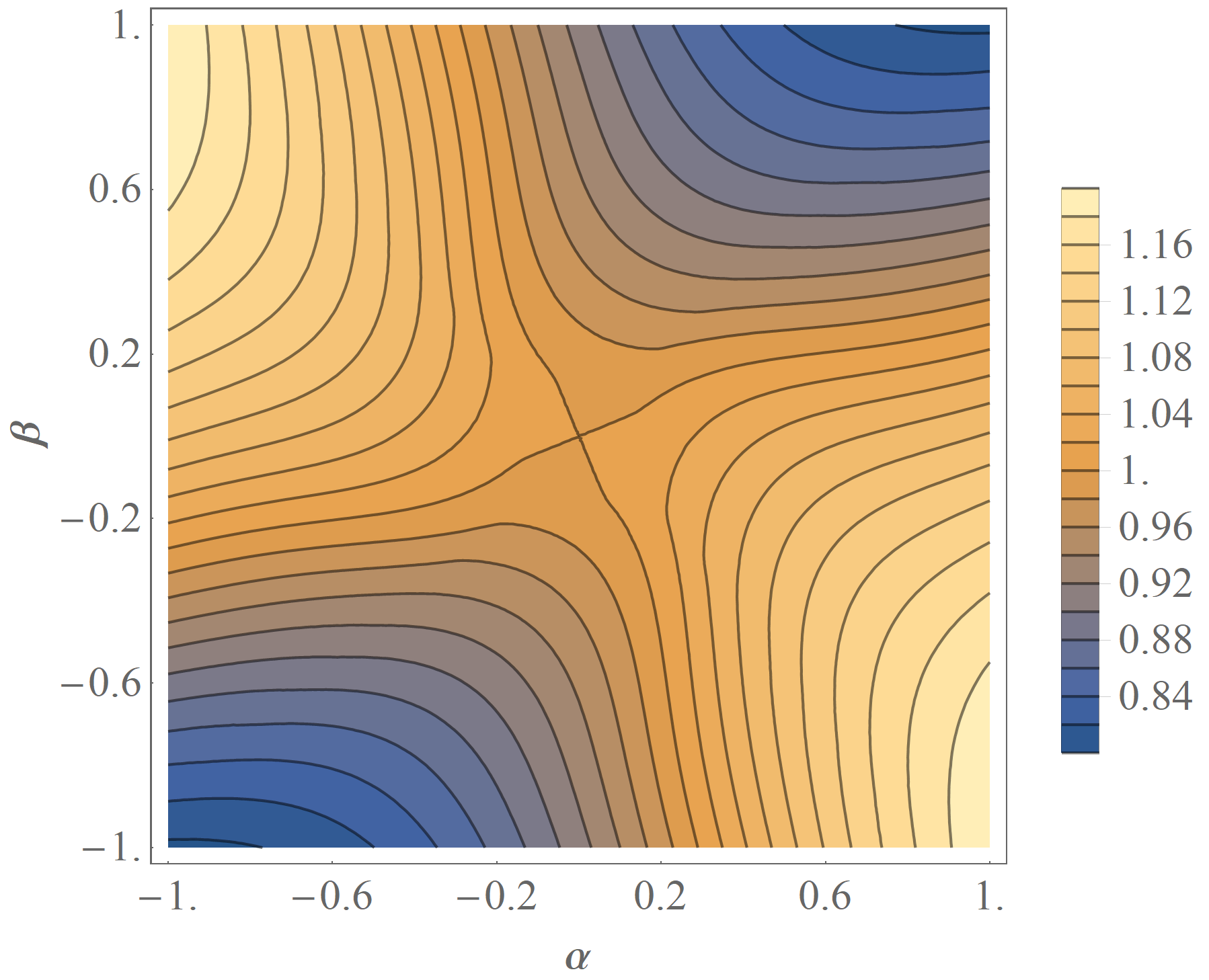} }  
	\caption{The upper, lower and Nash equilibrium state-value functions under $T=1$.}
\end{figure}

\begin{figure}[H]
	\centering
	\subfigure[\footnotesize{Optimal actual control inputs and game matrix at state $(0.2,-0.2)^\mathrm{T}$}]{\includegraphics[height=7cm]{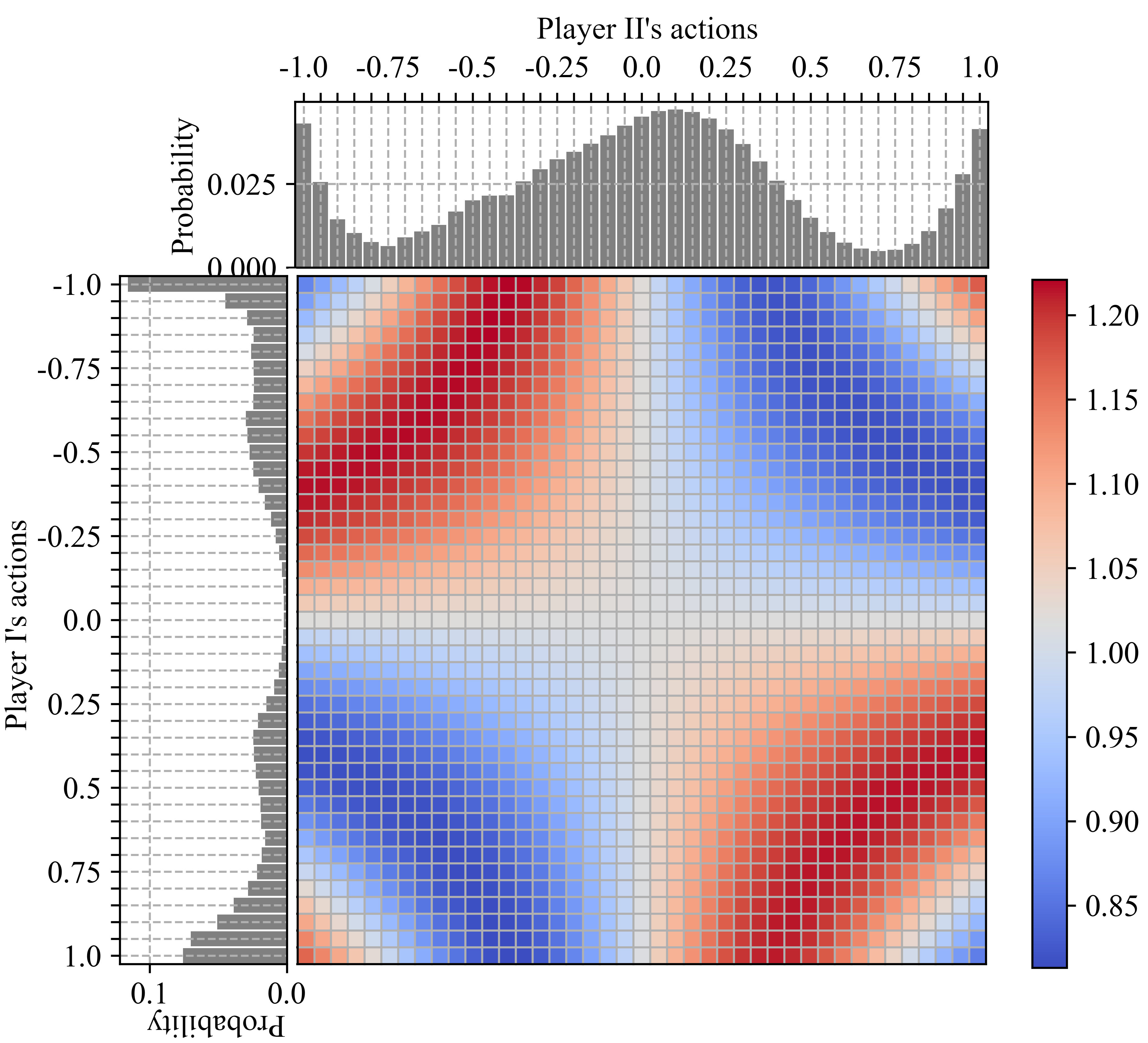} } 
	\subfigure[\footnotesize{Optimal actual control inputs and game matrix at state $(0.3,0.35)^\mathrm{T}$}]{\includegraphics[height=7cm]{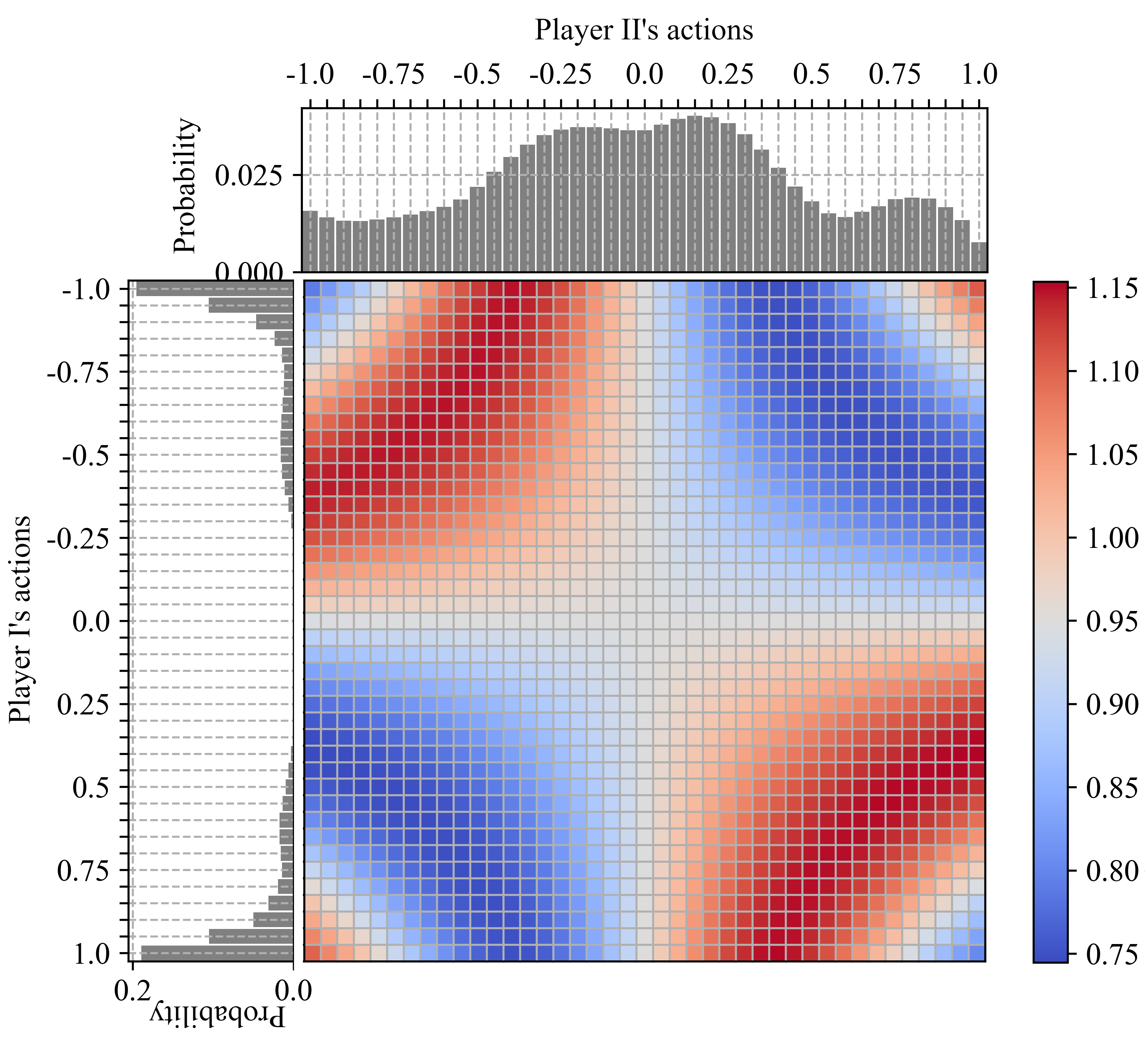} }  
	\caption{At state $(0.2,-0.2)^\mathrm{T}$ and $(0.3,0.35)^\mathrm{T}$, both players have 
	many actions that are played with positive probability. The values in different entries of the game matrix 
	are expressed as different colors.}
\end{figure}

\subsubsection{Battle between different strategies}
In this example, we also simulate some battles between different control policies. 
The settings are the same
as the ones in Example 1. 
The simulation results are shown in Table 2. 
Table 2 shows that, the values in different entries have significant difference. 
For player II, compared to other control policies, our method is a dominant control policy.  
That is, for player II, choosing our method always gives a better outcome than choosing other control policies, no matter what player I do.
For player I, the best control policy is our method when player II chooses our method as its control policy.

\begin{table}[h]
	\centering
	\caption{Battle results of Example 2.}
	\begin{tabular}{cc|c|c|c|}
	\cline{3-5}
											  &   & \multicolumn{3}{c|}{Player II's control policy} \\ \cline{3-5} 
											  &   & Min-max type      & Max-min type     & Our method    \\ \hline
	\multicolumn{1}{|l|}{\multirow{3}{*}{\rotatebox{90}{Player I's control policy}    }} & \rotatebox{90}{Min-max type} &    16.7026    &    15.23    &    7.5577   \\ \cline{2-5} 
	\multicolumn{1}{|l|}{}                    & \rotatebox{90}{Max-min type} &    10.4195    &    11.5609    &    8.1819   \\ \cline{2-5} 
	\multicolumn{1}{|l|}{}                    & \rotatebox{90}{Our method} &    11.6277    &    11.7377    &    8.5097  \\ \hline
	\end{tabular}
\end{table}

\section{Conclosions}
In this paper, a method
is developed to solve the rolling Nash equilibrium of zero-sum differential games. 
The first step is discretize time into several intervals with small size. Then,
with the aid of state-value function, the differential game 
is translated into a recursion consisting of normal-form game, and based on the action abstraction and regret matching 
the Nash equilibrium of each normal-form game can be obtained. 
When use our method to deal with a on-line control problem, the state-value function can be stored in memory 
to improve the real-time property.
This method is effective for both the situations that the saddle point exists or
does not exist. The analysis of existence of the
saddle point are avoided. For the situation that the saddle point exists, 
our method can give the pure control policy. 
For the situation that the saddle point does not exist, 
our method can give the probability distribution of the mixed control policy.
In order to improve the accuracy and reduce the grid quantity,
possible future developments will address a more advanced interpolation method instead of the linear
interpolation described here.

\section*{Acknowledgements}
\par The authors gratefully acknowledge support from National Defense Outstanding Youth Science Foundation (Grant No. 2018-JCJQ-ZQ-053), 
and Central University Basic Scientific Research Operating Expenses Special Fund Project Support (Grant No. NF2018001).
Also, the authors would like to thank
the anonymous reviewers, associate editor, and editor for
their valuable and constructive comments and suggestions.

\section*{Appendix}
\par In this section, we introduce the linear interpolation applied in this paper. 
Let $x_0$ denote a system state, and $x_1$, $x_2$, $x_3$ are the three nearest grid points 
around $x_0$. These three grid points can constitute a right triangle cell.
The coordinate of $x_i$ is denoted as $(\alpha_i,\beta_i) $.
According to the nearest grid point of $x_0$, the interpolation can be divided into four cases, see Fig. 9. 
Let $V_i$ denote the state-value of point $x_i$ (the state-value of the grid points are stored in the two-dimensional array $\mathcal{V}$).
The goal of interpolation is to estimate $V_0$.
\begin{figure}[H]
    \centering
    \subfigure[The nearest grid point is located at the bottom left.]{\includegraphics[height=3.5cm]{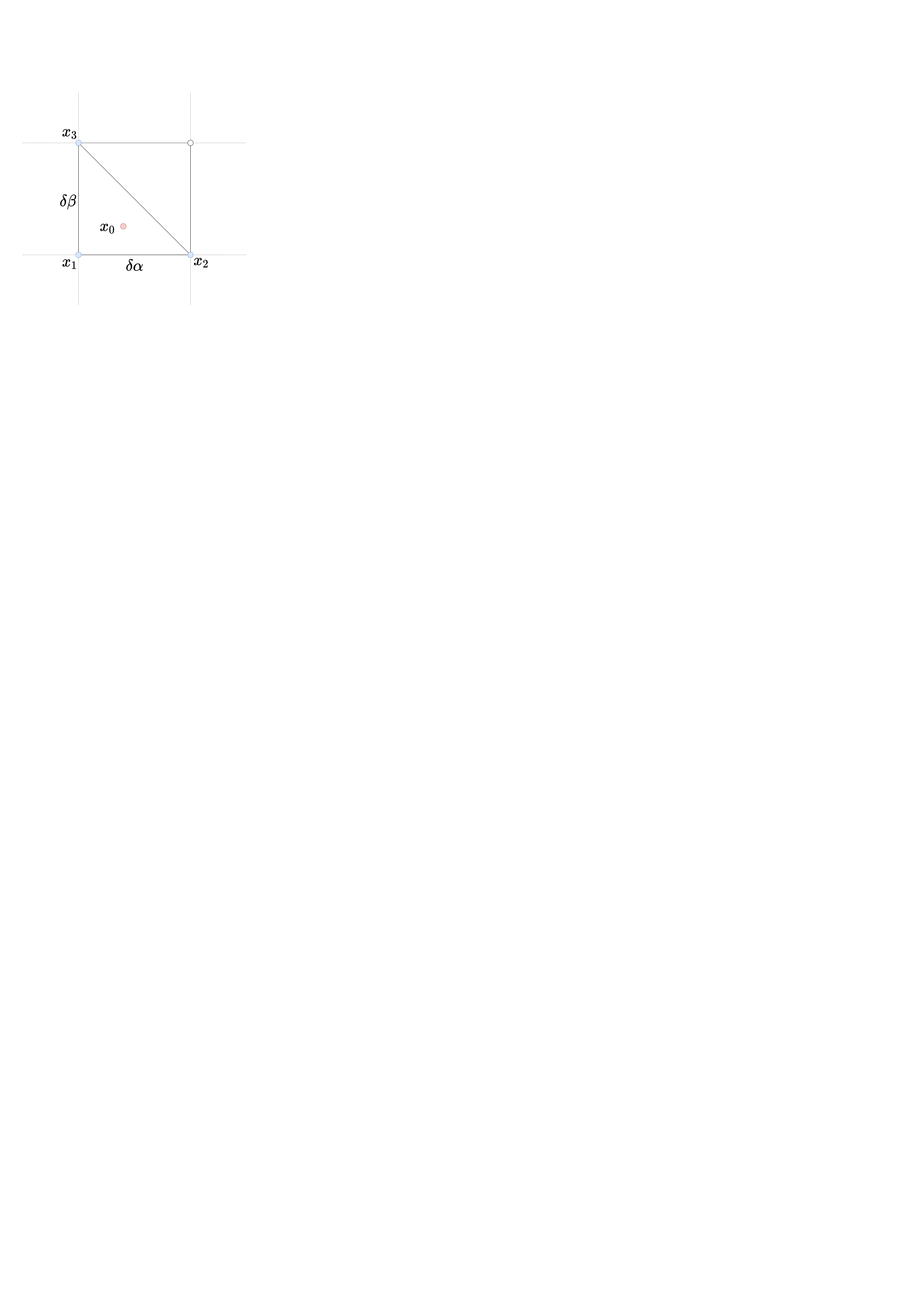} } \quad 
    \subfigure[The nearest grid point is located at the top left.]{\includegraphics[height=3.5cm]{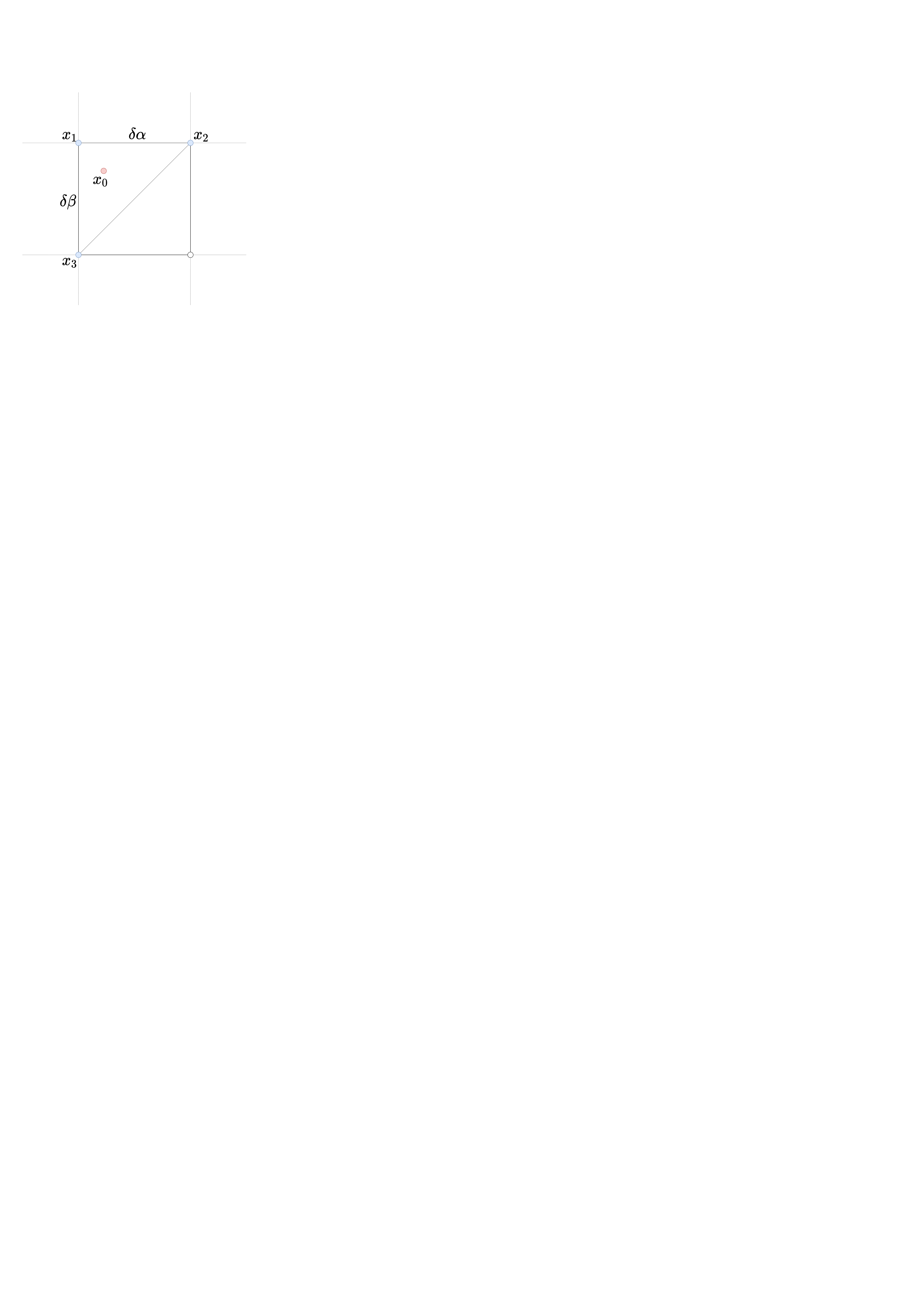}}\\
    \subfigure[The nearest grid point is located at the top right.]{\includegraphics[height=3.5cm]{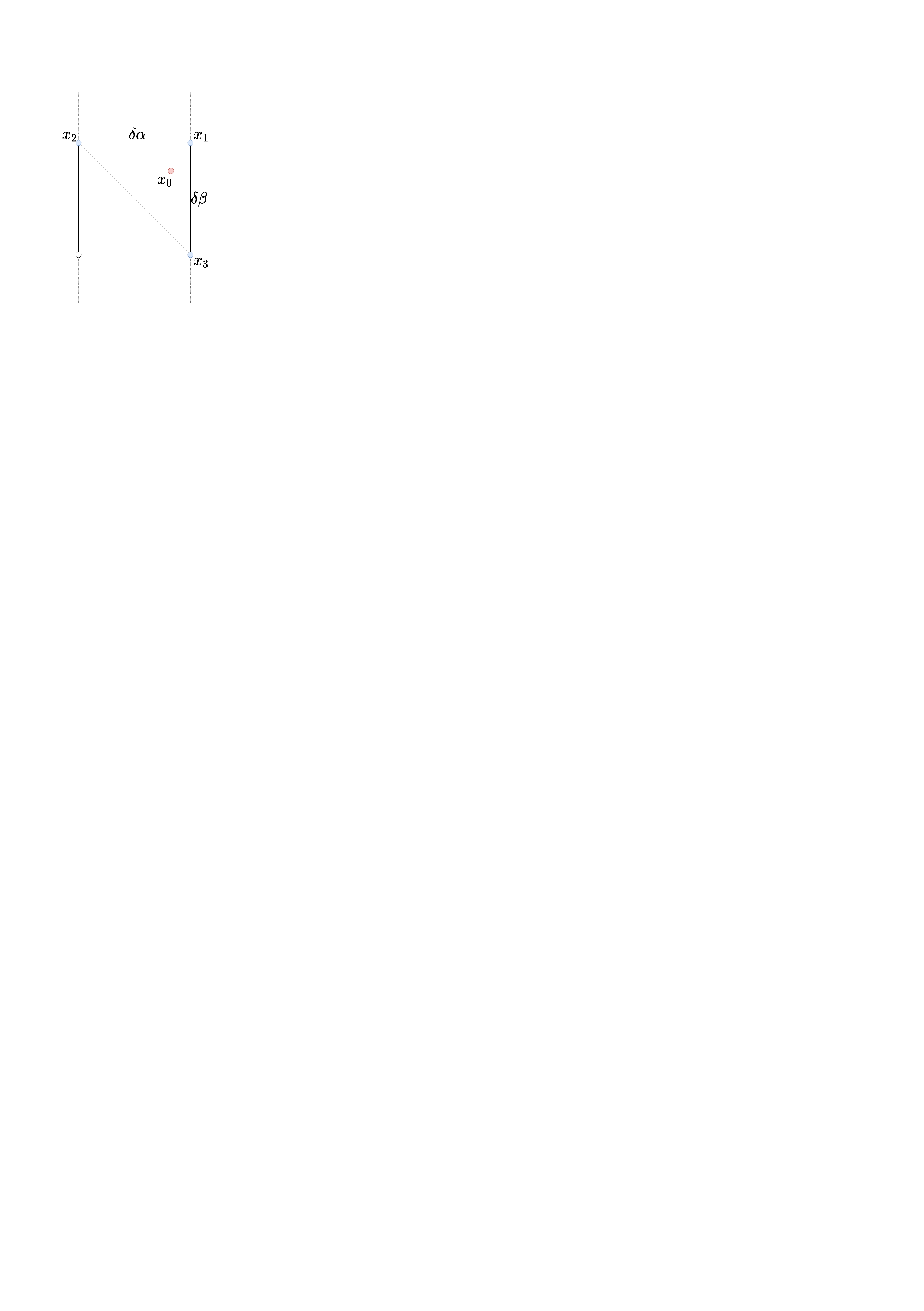}} \quad 
    \subfigure[The nearest grid point is located at the bottom right.]{\includegraphics[height=3.5cm]{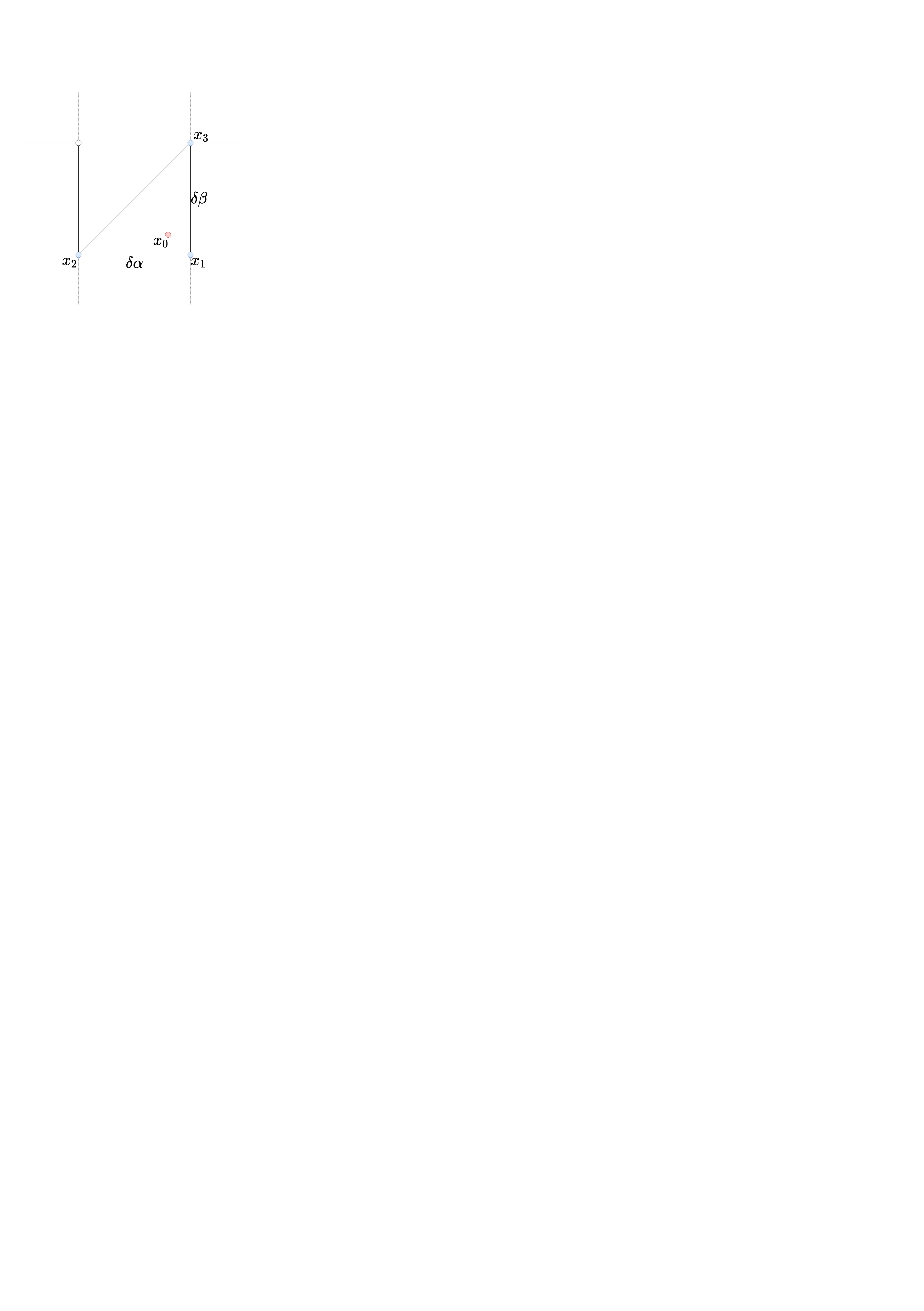}}
    \caption{Interpolation of different cases.}
\end{figure}
The explanations of the symbols in Fig. 9 are as following: 
\begin{align}
	\delta \alpha=\alpha_2-\alpha_1,\delta \beta=\beta_3-\beta_1
\end{align}
Then 
\begin{align}
	V_0=\frac{V_2-V_1}{\delta \alpha} (\alpha_0-\alpha_1)+\frac{V_3-V_1}{\delta \beta}(\beta_0-\beta_1)+V_1
\end{align}

% References

\bibliographystyle{unsrt}
% \bibliography{Bibliography/IEEEabrv,Bibliography/BIB_xx-TIE-xxxx}\ %IEEEabrv instead of IEEEfull
\bibliography{mybib}

\end{document}